\documentclass[prb,amsmath,amssymb,eqsecnum,twocolumn]{revtex4} 
\usepackage{graphicx}
\usepackage{amsmath}
\usepackage{amssymb}
\usepackage{euscript}
\usepackage{epsfig}
\usepackage{rotating}
\usepackage{color}
\usepackage{hhline}
\usepackage{hyperref}
\newcommand{\arcsinh}{\operatorname{arcsinh}}
\newcommand{\R}{{\mathbb{R}}}
\renewcommand{\Im}{\operatorname{Im}\, }

\newcommand{\lac}{La$_2$CuO$_4$}
\newcommand{\src}{Sr$_2$CuO$_2$Cl$_2$}
\newcommand{\kni}{K$_2$NiF$_4$}
\newcommand{\lan}{La$_2$NiO$_4$}
\newcommand{\kfe}{KFeF$_4$}
\newcommand{\rbm}{Rb$_2$MnF$_4$}

\begin{document} 
\title{
Scaling relations in the quasi-two-dimensional Heisenberg antiferromagnet
      }
\author{Antoine Praz$^1$, Christopher Mudry$^1$ and M.~B. Hastings$^2$} 

\affiliation{
$^1$Condensed Matter Theory Group, Paul Scherrer Institut, 
CH-5232 Villigen PSI, Switzerland
            }
\affiliation{
$^2$Theoretical Division, Los Alamos National Laboratory, 
Los Alamos, New Mexico 87545, USA
            }

\date{\today} 


\begin{abstract}
The large-$N$ expansion of the quasi-two-dimensional
quantum nonlinear $\sigma$ model is used in order to establish 
experimentally applicable
universal scaling relations for the quasi-two-dimensional Heisenberg 
antiferromagnet. We show that, at $N=\infty$, 
the renormalized coordination number introduced by 
Yasuda \textit{et al.}
[Phys.~Rev.~Lett.~\textbf{94}, 217201 (2005)]
is a universal number in the limit of $J'/J\rightarrow 0$.
Moreover, similar scaling relations proposed by
Hastings and Mudry
[Phys. Rev. Lett.~\textbf{96}, 027215 (2006)]
are derived at $N=\infty$ for the three-dimensional static spin susceptibility
at the wave vector $(\pi,\pi,0)$, 
as well as for the instantaneous structure factor at the same wave vector.
We then use $1/N$ corrections to study the relation between
interplane coupling, correlation length, and critical temperature, and show
that the universal scaling relations lead to logarithmic corrections
to previous mean-field results.
\end{abstract}

\maketitle 

\section{
Introduction
        }
\label{sec: Introduction} 

Many materials, such as copper oxides and other layered perovskites,
are known to be nearly two-dimensional magnets.
While in certain intermediate temperature ranges these systems are
well described by purely two-dimensional models, 
three dimensionality is restored at temperatures
below an energy scale that is governed by the ratio between 
the interlayer coupling $J'$ and the intraplanar exchange parameter $J$.
For example, two-dimensional quantum Heisenberg antiferromagnets (AFs) 
do not support long-range collinear magnetic order at any finite temperature
according to the Mermin-Wagner theorem,\cite{Mermin-Wagner} 
while real layered systems, such as La$^{\ }_{2}$CuO$^{\ }_4$, do. 
The anisotropy $J'/J$ of a quasi-two-dimensional material 
can be determined from the spin-wave dispersion below the ordering
temperature. It can also be determined from the measured ordering temperature 
provided one understands the two-dimensional to three-dimensional 
crossover that manifests itself in the
dependence on $J'/J$ of the ordering temperature.

A common approximation for the ordering temperature 
$T^{\ }_{\mathrm{c}}$ 
of a quasi-$n$-dimensional magnetic system
is the random-phase approximation (RPA),~\cite{Scalapino,Schultz96}
which predicts that
\begin{eqnarray}
J'\chi^{(n)}_{\mathrm{c}}=
\frac{1}{z^{(n)}}.
\label{eq: def RPA}
\end{eqnarray}
Here $\chi^{(n)}_{\mathrm{c}}$ is the exact static susceptibility
associated to the magnetic order for the underlying
$n$-dimensional subsystem evaluated 
at the ordering temperature $T^{\ }_{\mathrm{c}}$
and $z^{(n)}$ is the coordination number of the $n$-dimensional subsystem.
Yasuda {\it et al.}~in Ref.~\onlinecite{Yasuda05}
have quantified the accuracy of the RPA in two steps. First, they computed
the three-dimensional AF ordering temperature $T^{\ }_{AF}$ with the help of
a quantum (classical) Monte Carlo (MC) simulation of 
a spin-$S$ ($S=\infty$ in the classical system) 
nearest-neighbor Heisenberg model on a cubic lattice
with AF exchange coupling $J'$ along the vertical axis and AF exchange coupling
$J$ within each layer of the cubic lattice.
Second, they computed the two-dimensional static staggered susceptibility 
$\chi^{(2)}_{s}$
evaluated at $T^{\ }_{AF}$ from step 1 after switching off $J'$.
They thus showed that, for small $J'/J$, 
$T^{\ }_{AF}$ is given by a modified random-phase approximation, 
in which the coordination number gets renormalized,
\begin{eqnarray}
z^{(2)}\longrightarrow\zeta^{(2)}.
\label{eq: def zeta}
\end{eqnarray} 
It turns out that the renormalization $\zeta^{(2)}$ 
of the coordination number $z^{(2)}=2$ 
converges as $J'/J\to0$ to a value that is independent of the
spin quantum number $S$ taking values in $1/2,1,\dots,\infty$.
This fact motivated them to conjecture the 
\textit{universality of the effective coordination number $\zeta^{(2)}$} 
in the limit $J'/J\rightarrow 0$.

The results of Yasuda \textit{et al}.~were shown by Hastings and Mudry in 
Ref.~\onlinecite{Hastings06}
to reflect the so-called renormalized classical (RC) regime 
of the underlying two-dimensional subsystem. Hastings and Mudry
predicted that if the two-dimensional subsystem
is characterized by a quantum critical (QC) regime, 
then the effective coordination number in the limit $J'/J\rightarrow 0$
is a \textit{universal function} of the ratio $c\beta^{\ }_{AF}/\xi^{(2)}$
-- a number of order 1 in the QC regime
as opposed to a vanishing number
in the RC regime --
where $c$ is the two-dimensional spin-wave velocity, 
$\beta^{\ }_{AF}=1/T^{\ }_{AF}$,
and $\xi^{(2)}$ is the two-dimensional correlation length.

Hastings and Mudry also proposed universal scaling relations involving 
observables of the quasi-two-dimensional system only. 
One of these scaling relations is obtained 
from multiplying the static three-dimensional spin susceptibility 
$\chi^{(3)}_{s}$ evaluated at the wave vector $(\pi,\pi,0)$ 
and at the N\'eel temperature with $J'$. 
Another scaling relation can be derived by multiplying 
the instantaneous structure 
factor at the transition temperature $S^{(3)}(\pi,\pi,0;t=0;T^{\ }_{AF})$ 
with $J'$ and the inverse N\'eel temperature $1/T^{\ }_{AF}$.
This last universal relation has the advantage of being directly measurable
with the help of inelastic neutron scattering.

The aim of this work is to verify the universal scaling laws proposed above
within the $N=\infty$ approximation to the quasi-two-dimensional
quantum nonlinear $\sigma$ model (QNLSM). 
Our starting point is the two-dimensional QNLSM
(a two-space and one-time NLSM),
which is believed to capture the physics of two-dimensional quantum Heisenberg 
AF at low energies.~\cite{Chakravarty88,Chakravarty89,Chubukov94}
Assuming that the physics of a single layer can be approximated by the
two-dimensional QNLSM, we introduce the interlayer coupling following 
Refs.~\onlinecite{Chakravarty88},
\onlinecite{Chakravarty89}, 
\onlinecite{Affleck96}, 
and~\onlinecite{Irkhin97}.
We then show that, in the $N=\infty$ approximation, 
the quantities
$J'\chi^{(2)}_{s}(T^{\ }_{AF})$, 
$J'\chi^{(3)}  (\pi,\pi,0;\omega=0,T^{\ }_{AF})$,
and 
$J'S^{(3)}(\pi,\pi,0;t=0;T^{\ }_{AF})/T^{\ }_{AF}$
each converge to universal scaling functions of
the dimensionless ratio between the thermal de Broglie wavelength
of spin waves and a correlation length
in the limit $J'/J\to 0$. As in the strict two-dimensional limit, 
different regimes can be distinguished depending on the value 
taken by this dimensionless ratio. 
The form of the nonuniversal corrections to the universal 
functions obtained in the limit $J'/J\rightarrow 0$ strongly 
depends on these regimes. In the renormalized classical (RC) regime,
which is dominated by classical thermal fluctuations, 
the nonuniversal corrections are relatively small for small $J'/J$, 
as is seen from MC simulations.~\cite{Yasuda05}
Near the critical coupling of the two-dimensional QNLSM,
fluctuation physics is predominantly quantum.
In this quantum regime, the universal constants obtained in the limit 
$J'/J\rightarrow 0$ depend on the way this limit is taken, i.e.,
on the value taken by the \textit{fixed ratio} (\ref{eq: def R-+}) 
between the characteristic length scale 
of the quantum fluctuations 
and the characteristic length scale at which interplane
interactions become important.
When this ratio is a large number, 
the fluctuation regime is the quantum critical (QC) regime. 
When this ratio is of order 1,
the fluctuation regime is the quantum disordered (QD) regime. 
When this ratio is much smaller than 1, the quasi-two-dimensional system
remains disordered all the way down to zero temperature.

The universal relations described above have recently been tested
numerically by Yao and Sandvik.~\cite{Yao06}
They have performed quantum Monte-Carlo simulations of spin-$1/2$ 
quasi-two-dimensional systems in the RC, QC, and QD regimes.
Their results qualitatively agree with the predictions 
of the large-$N$ expansion proposed here.

We state the universal scaling laws in Sec.~\ref{sec: Universal scaling relations}.
The quasi-two-dimensional QNLSM, which is the basis of our analysis, is defined in
Sec.~\ref{sec: Quasi-two-dimensional QNLSM}. The $N=\infty$ approximation to the
N\'eel temperature is given in Sec.~\ref{sec: Neel T in the N=infty limit}.
Scaling laws in the $N=\infty$ approximation are derived for the quasi-two-dimensional
QNLSM in 
Sec.~\ref{sec: Scaling laws in the N=infty limit}
and for its classical limit, the  quasi-two-dimensional NLSM, in
Sec.~\ref{sec: Scaling laws in the N=S=infty limit}.
Conclusions are presented in Sec.~\ref{sec: Conclusions}
while we defer to the appendixes for the derivation of the counterparts 
to these scaling laws in the case of the quasi-one-dimensional Ising model.
Among the most important conclusions is an extension to finite $N$, where
we relate $J',\, J,\,  T_{AF},\, \xi^{(2d)}$ and find scaling relations that differ
from mean-field estimates by logarithmic corrections.

\section{
Universal scaling relations
        }
\label{sec: Universal scaling relations}

We are going to formulate precisely the universal relations discussed 
in Sec.~\ref{sec: Introduction}. 
It is instructive and necessary to first identify the relevant length scales 
of the system near the crossover between 
two dimensionality
and 
three dimensionality. These will be used to construct the relevant
dimensionless scaling variables whose values will allow us, in turn,  
to separate different scaling regimes.

\subsection{
Characteristic length scales
           }
\label{subsec: Characteristic length scales}

We start with the Hamiltonian describing a spatially anisotropic
Heisenberg AF on a cubic lattice,
\begin{equation}
H:= 
J
\sum_{\langle ii'\rangle}
\sum_{j} 
\boldsymbol{S}^{\ }_{i,j} 
\cdot 
\boldsymbol{S}^{\ }_{i',j}
+ 
J'
\sum_{i}
\sum_{\langle jj'\rangle}
\boldsymbol{S}^{\ }_{i,j}
\cdot
\boldsymbol{S}^{\ }_{i,j'}.
\label{eq: micro Hamiltonian}
\end{equation}
Planar coordinates of the cubic lattice sites are denoted by the letters
$i$ and $i'$. The letters $j$ and $j'$ label each plane.
The angular bracket $\langle ii'\rangle$ denotes
planar nearest-neighbor sites.
The angular bracket $\langle jj'\rangle$ denotes two consecutive planes.
A spin operator located at site $i$ in the $j$th plane is denoted by
$\boldsymbol{S}^{\ }_{i,j}$. 
It carries the spin quantum number $S=1/2,1,\dots$ .
The coupling constants are AF, $J,\, J'>0$,
and are strongly anisotropic, $J\gg J'$.

We are interested in the thermodynamic properties of the system 
at the AF ordering temperature $T^{\ }_{AF}$. 
At this temperature, there exists a diverging length scale, 
the temperature-dependent correlation length 
\begin{eqnarray}
\xi\equiv
\xi^{(3)}_{\pi,\pi,\pi},
\end{eqnarray}
for AF correlations at the wave vector $(\pi,\pi,\pi)$.
For a quasi-two-dimensional AF, 
the strong spatial anisotropy $J\gg J'>0$
allows us to identify the characteristic length scale
\begin{subequations}
\begin{equation}
\xi':= 
\sqrt{\frac{1}{\mathcal{Z}^{\prime}}}\sqrt{\frac{J}{J'}}\
a,
\label{eq: definition of xi prime}
\end{equation}
where $a$ is the lattice spacing of the cubic lattice
and $\mathcal{Z}^{\prime}$ is a multiplicative renormalization.%
\cite{footnote: Z'}
In the $N=\infty$ approximation considered in this paper,
$\left.\mathcal{Z}^{\prime}\right|^{\ }_{N=\infty}=1$. We will omit 
this renormalization factor in the following until we treat
$1/N$ corrections in 
Sec.~\ref{subsec: Experimental discussion of the first scaling law}.

In a good quasi-two-dimensional Heisenberg AF 
\begin{eqnarray}
0<J'\ll J
\Longrightarrow
\xi'\gg a.
\end{eqnarray}
\end{subequations}
It was proposed~\cite{Affleck96,Irkhin97} that $\xi'$ 
determines the crossover between two and three dimensionality.
Heuristically, the temperature dependence of the
AF correlation length is nothing but that of 
the two-dimensional underlying system $\xi^{(2)}$
as long as $\xi^{(2)}\ll\xi'$. Upon approaching $T^{\ }_{AF}$
from above, $\xi^{(2)}$ grows until 
$\xi^{(2)}\sim\xi'$. Below this temperature,
three-dimensionality is effectively recovered 
and AF long-range order becomes possible.

In the regime of temperature above the ordering temperature
$T^{\ }_{AF}$ for which $\xi^{(2)}\gg a$, 
the two-dimensional spin-wave velocity $c$ is well defined
and depends weakly on temperature.
Together with the inverse N\'eel temperature $\beta^{\ }_{AF}$, 
we can then build the thermal de Broglie wavelength $c\beta^{\ }_{AF}$.
Equipped with the characteristic length scales $c\beta^{\ }_{AF}$
and $\xi'$, we can define three regimes. 
(i) The \textit{renormalized classical} (RC) regime 
is defined by the condition
\begin{eqnarray}
a\ll c\beta^{\ }_{AF}\ll\xi'.
\label{eq: regime i}
\end{eqnarray}
(ii) The \textit{quantum critical} (QC) regime 
is defined by the condition 
\begin{eqnarray}
a\ll c\beta^{\ }_{AF}\simeq\xi'.
\label{eq: regime ii}
\end{eqnarray}
(iii)  The \textit{quantum disordered} (QD) regime 
is defined by the condition 
\begin{eqnarray}
a\ll\xi'\ll c\beta^{\ }_{AF}.
\label{eq: regime iii}
\end{eqnarray}

In the first regime (\ref{eq: regime i}),
fluctuations are predominantly two-dimensional and thermal.
The planar correlation length diverges exponentially fast with
decreasing temperature. This leads to a rather sharp crossover 
between two- and three-dimensionality. As a corollary, there will be
small nonuniversal corrections to universality for finite $J'/J$.
In the second, Eq.~(\ref{eq: regime ii}), 
and third, Eq.~(\ref{eq: regime iii}), 
regimes, fluctuations are predominantly two-dimensional and quantum.
In the QC regime (\ref{eq: regime ii}), 
the underlying two-dimensional fluctuations are quantum critical.
In the QD regime (\ref{eq: regime iii}), they are quantum disordered.

Common to all three regimes 
(\ref{eq: regime i})-(\ref{eq: regime iii})
is the fact that the lattice spacing is much smaller than the
de Broglie wavelength constructed from the two-dimensional spin-wave
velocity and the three-dimensional ordering temperature.
It is then reasonable to expect that
planar fluctuations of the microscopic Hamiltonian%
~(\ref{eq: micro Hamiltonian})
can be captured by an effective low-energy and long-wavelength 
effective continuum theory. 
We choose this effective field theory to be the 
two-dimensional QNLSM.~\cite{Chakravarty88,Chakravarty89,Chubukov94}
To account for interplanar fluctuations, we preserve the lattice
structure by coupling in a discrete fashion an infinite
array of two-space and one-time QNLSM. We shall call this effective theory 
the quasi-two-dimensional QNLSM.
The ratio $c\beta^{\ }_{AF}/\xi'$ is then determined by the parameters of 
the quasi-two-dimensional QNLSM, i.e., by the two-dimensional 
spin-wave velocity $c$
and spin stiffness $\rho^{\ }_{s}$
(or the gap $\Delta$), as well as on $J'/J$.
These are related to the microscopic parameters of the quantum Heisenberg AF.
While the model with nearest-neighbor couplings 
and with physical spins ($S\ge 1/2$)
is known to be in the RC regime at low temperatures, 
the addition of terms in the Hamiltonian%
~(\ref{eq: micro Hamiltonian})
such as frustrating next-nearest-neighbor couplings, say,
allows us to realize the QC or QD regimes. 
We thus consider the two-dimensional 
spin-wave velocity $c$
and spin stiffness $\rho^{\ }_{s}$ 
(or the gap $\Delta$) 
as phenomenological parameters.

The scaling functions we are looking for will depend 
on the ratios of the lengths $\xi'$, $c\beta_{AF}$, and, 
if we are interested in an observable of the two-dimensional underlying system, 
of $\xi^{(2)}(\beta_{AF})$. 
However, at the transition temperature and in the limit of $J'/J\to 0$, 
only one of the two ratios is 
independent, and the scaling functions will indeed depend only on 
one dimensionless parameter.\cite{Hastings06}

\subsection{
Universal scaling functions
           }
\label{subsec: scaling functions}

The first universal scaling function, 
suggested in Ref.~\onlinecite{Hastings06},
involves the two-dimensional static and staggered spin susceptibility,
\begin{subequations}
\label{eq: claim for F1}
\begin{eqnarray}
\lim_{J'/J\to0}
J'\chi^{(2)}_{s}(\beta^{\ }_{AF})= 
F^{\ }_1 
\big( 
c\beta^{\ }_{AF}/\xi^{(2)}(\beta^{\ }_{AF}) 
\big),
\label{eq: first scaling law}
\end{eqnarray}
where
\begin{eqnarray}
\chi^{(2)}_{s}(\beta^{\ }_{AF}):= 
\chi^{(2)}(\boldsymbol{k}^{\ }_{AF},\omega=0;\beta^{\ }_{AF})
\label{eq: def 2-dim iso  static spin susc}
\end{eqnarray}
\end{subequations}
with 
$\boldsymbol{k}^{\ }_{AF}=(\pi,\pi)$. 
In the RC regime, 
\begin{eqnarray}
F^{\ }_1 
\big( 
c\beta^{\ }_{AF}/\xi^{(2)}(\beta^{\ }_{AF}) 
\big)
=
F^{\ }_1(0)
\equiv1/\zeta^{(2)}
.
\end{eqnarray}
MC simulations give $\zeta^{(2)}\simeq 1.3$,~\cite{Yasuda05}
while the RPA approximation leads to $\zeta^{(2)}=2$. 
Using the $N=\infty$ approximation to the quasi-two-dimensional QNLSM, 
we will find $\zeta^{(2)}=1$ in the RC regime.

The two-dimensional spin-susceptibility is
inaccessible experimentally at the three-dimensional ordering temperature. 
In order to mimic the two-dimensional AF wave vector 
in the three-dimensional system, 
we choose to work at the wave vector
$(\pi,\pi,0)$.
We claim the existence of the universal scaling function 
\begin{subequations}
\label{eq: claim for F2}
\begin{equation}
\lim_{J'/J\to0}
J'\chi^{(3)}_{s}\big(\beta^{\ }_{AF}\big)=
F^{\ }_2\big( c\beta^{\ }_{AF} /\xi' \big),
\end{equation}
where we have defined
\begin{equation}
\chi^{(3)}_{s}\big(\beta^{\ }_{AF}\big):=
\chi^{(3)}\big(\pi,\pi,0,\omega=0;\beta^{\ }_{AF}\big).
\end{equation}
\end{subequations}
Observe that any vector of the form 
$(\pi,\pi,k^{\ }_z)$ with $k^{\ }_z\neq \pi$ 
would lead to the same conclusion but for different universal scaling functions that
depend on $k^{\ }_{z}$. 
In the $N=\infty$ approximation of the quasi-two-dimensional QNLSM, 
we find that $F^{\ }_{2}(x)=1/4$
takes the same value in the RC, QC, and QD regimes. 
This suggests that this scaling relation is rather robust 
and well suited for numerical studies.

Finally, we claim the existence of the universal scaling function
\begin{subequations}
\label{eq: claim for F3}
\begin{equation}
\lim_{J'/J\to0}
\beta^{\ }_{AF} 
J'S^{(3)}\big(\pi,\pi,0,t=0;\beta^{\ }_{AF}\big)=
F^{\ }_3\big( c\beta^{\ }_{AF} /\xi' \big),
\end{equation}
where $S^{(3)}$ is the instantaneous structure factor
\begin{equation}
S^{(3)}(\boldsymbol{q},q_z,t=0;\beta):=
-
\int_{-\infty}^{+\infty}\frac{d\omega}{\pi}
\frac{\Im\chi^{(3)}(\boldsymbol{q},q_z,\omega;\beta)}{e^{\beta\omega}-1}.
\label{eq: definition of S}
\end{equation}
\end{subequations}
In the $N=\infty$ approximation of the quasi-two-dimensional QNLSM, 
we find that $F^{\ }_{3}(x)=(x/4)\coth x$.

The reminder of the paper is devoted to proving the existence of these
three universal scaling functions and to their computation 
in the $N=\infty$ approximation of the quasi-two-dimensional QNLSM.

\section{
Quasi-two-dimensional QNLSM
        }
\label{sec: Quasi-two-dimensional QNLSM}

The quantum nonlinear $\sigma$ model (QNLSM) 
was successfully used to study the low-energy 
and long-wavelength physics of the
quantum one-dimensional Heisenberg AF,~\cite{Haldane83} the 
quantum two-dimensional Heisenberg model AF,~\cite{Chakravarty88,Chakravarty89,Chubukov94}
and the quantum quasi-two-dimensional Heisenberg AF.~\cite{Affleck96,Irkhin97}
In Haldane's mapping of the quantum Heisenberg AF to the QNLSM,%
~\cite{Haldane83}  
a crucial role is played by the correlation length.
A large correlation length gives the possibility to 
separate slow from fast modes. Integration over fast modes can 
then be carried out perturbatively, retaining only slow modes.
In the quasi-two-dimensional quantum Heisenberg AF, 
the length scale $\xi'$ defined in Eq.~(\ref{eq: definition of xi prime})
can also be used as a characteristic length scale
to separate the fast from the slow modes in Haldane's mapping.
These fast modes can then be integrated out 
following the procedure used by Haldane,~\cite{Haldane83}
and the partition function can be expressed in terms 
of a path integral over unit vectors. 
In the isotropic limit, the three-dimensional QNLSM, 
which is a pure field theory, is recovered. 
However, as noticed in 
Refs.\ \onlinecite{Chakravarty88,Chakravarty89},\ \onlinecite{Affleck96},
and\ \onlinecite{Irkhin97}, 
the large values of $\xi'$ allows us to take the continuum limit only 
within the planes. The resulting partition function has the form
\begin{subequations}
\label{eq: def d-dim QNLSM}
\begin{equation}
Z:= 
\int
\left\{
\prod_{\boldsymbol{x},{\tau},j}
\mathcal{D}[\boldsymbol{\sigma}^{\ }_j(\boldsymbol{x},\tau)]\,
\delta\big(\boldsymbol{\sigma}^2_j(\boldsymbol{x},\tau)-1\big)
\right\} 
e^{-\mathcal{S}[\boldsymbol{\sigma}]},
\label{eq: def quasi-2-dim QNLSM}
\end{equation}
where $\boldsymbol{\sigma}^{\ }_j(\boldsymbol{x},\tau)\in \R^N$, 
the integer $N=3$ for real spin systems,
and $\tau$ is the imaginary time.
The action in Eq.~(\ref{eq: def quasi-2-dim QNLSM})
can be divided in two parts, 
\begin{eqnarray}
S=S^{\ }_{0}+S^{\ }_{1},
\label{eq: S = S0+S1}
\end{eqnarray}
where 
\begin{equation}
S^{\ }_0:= 
\frac{\rho^{\ }_{s}}{2}
\sum_{j}
\int\limits_0^{\beta}d\tau 
\int d^2\boldsymbol{x}\, 
\left[
c^{-2} 
\left(
\partial^{\ }_{\tau}\boldsymbol{\sigma}^{\ }_j
\right)^2
+
\left(
\boldsymbol{\nabla}\boldsymbol{\sigma}^{\ }_j
\right)^2
\right]
\label{eq: S = S0}
\end{equation}
is the action of the two-dimensional QNLSM 
on a collection of independent planes labeled by $j$
and 
\begin{equation}
S^{\ }_{1}:= 
\frac{\rho^{\ }_{s}}{2}
\frac{J'}{J}
\sum_{j}
\int\limits_0^{\beta}d\tau 
\int d^2\boldsymbol{x}\, 
\boldsymbol{\sigma}^{\ }_j\cdot\boldsymbol{\sigma}^{\ }_{j+1}
\label{eq: S = S1}
\end{equation}
\end{subequations}
describes the interplane coupling.\cite{footnote: nnn nnnn etc}
The bare planar spin stiffness 
$\rho^{\ }_{s}=JS^2$ 
and the bare planar spin-wave velocity
$c=2\sqrt{2}SJ$
provided units such that $a=1$ have been chosen.
The $O(N)$ model is constructed allowing $N$ 
to take any value larger than 2 in $\mathbb{N}$. 
The local constraint $\boldsymbol{\sigma}^2_j(\boldsymbol{x},\tau)=1$
can be ensured by a Lagrange multiplier 
$\lambda^{\ }_j(\boldsymbol{x},\tau)$. 
The $N=\infty$ approximation is then obtained 
after integrating out the original
fields $\boldsymbol{\sigma}^{\ }_j$ and by
expressing the original partition function in the form 
\begin{subequations}
\label{eq: large-N partition function}
\begin{equation}
Z= 
\int 
\left\{
\prod_{\boldsymbol{x},{\tau},j}
\mathcal{D}[\lambda^{\ }_{j}(\boldsymbol{x},\tau)]
\right\} 
e^{-NS^{\ }_{\mathrm{eff}}[\lambda]},
\end{equation}
where the effective action is now
\begin{equation}
\begin{split}
S^{\ }_{\mathrm{eff}}[\lambda]=&
-
\frac{c}{2g}
\sum_{j}\int\limits_0^\beta d\tau \int d^{2}\boldsymbol{x}\, 
\lambda^{\ }_{j}(\boldsymbol{x},\tau)
\\
&
+
\frac{1}{2}
\mathrm{Tr}\,
\ln 
\left(
-
\Delta^{\ }_{\boldsymbol{x},\tau}
-
\alpha\Delta^{\ }_{z}
+
\lambda^{\ }_{j}(\boldsymbol{x},\tau)
\right).
\end{split}
\label{eq: effective action}
\end{equation}
We have introduced the bare coupling 
\begin{eqnarray}
g:=cN/\rho^{\ }_{s},
\label{eq: def g}
\end{eqnarray}
the bare spatial anisotropy strength 
\begin{eqnarray}
\alpha:=
J'/J=
(1/\xi')^{2}, 
\end{eqnarray}
(remember that $a\equiv 1$), and 
\begin{eqnarray}
\Delta^{\ }_{\boldsymbol{x},\tau}&:=& 
c^{-2}\partial^{2}_{\tau}
+
\boldsymbol{\nabla}^2,
\\
\Delta^{\ }_{z}\boldsymbol{\sigma}_j&:=& 
2\boldsymbol{\sigma}^{\ }_j
-
\boldsymbol{\sigma}^{\ }_{j+1}
-
\boldsymbol{\sigma}^{\ }_{j-1}.
\end{eqnarray}
\end{subequations}
The parameter $N$ enters explicitly the action as a prefactor only. 
In the limit of large-$N$, any observable can be expanded 
in powers of $1/N$ with the leading order corresponding
to the saddle-point approximation. 
[for a review, see Ref.~\onlinecite{Polyakov87}].
The two-point spin correlation function is 
\begin{subequations}
\begin{eqnarray}
G^{ab}_{j,j'}(x,x')&=&
\delta^{ab}
Z^{-1}
\int 
\mathcal{D}[\lambda]
e^{-NS^{\ }_{\mathrm{eff}}[\lambda]} 
\widehat{G}^{\ }_{j,j'}(x,x';\lambda),
\nonumber\\
&&
\label{eq: 2-point spin corrleation fct}
\end{eqnarray}
with
\begin{eqnarray}
\widehat{G}^{\ }_{j,j'}(x,x';\lambda)&:=&
\left\langle 
x,j
\left|
\frac{
1
     }
     {
-
\Delta^{\ }_{\boldsymbol{x},\tau}
-
\alpha\Delta^{\ }_z
+
\lambda\delta^{\ }_{jj'} 
     }
\right| 
x',j'
\right\rangle, 
\nonumber\\
&&
\label{eq: two-point Green function}
\end{eqnarray}
\end{subequations} 
to leading order in the large-$N$ expansion.
We have introduced $x\equiv(\boldsymbol{x},c\tau)$. 
Because we approach the N\'eel ordering temperature from above,
we can assume isotropy in spin space of the spin correlation functions
and drop the spin indices on both sides of
Eq.~(\ref{eq: 2-point spin corrleation fct}).
The spin-isotropic susceptibility is then related to the 
correlation function in momentum space,~\cite{footnote: Fourier conv}
\begin{equation}
G(\boldsymbol{k},k^{\ }_z,\omega):=
\sum_{j}
\int_x
e^{
-\mathrm{i}
\left(
kx
+
k^{\ }_{z}j
\right)
  }
G^{\ }_{j,0}(x,0)
\label{eq: def Fourier trsf}
\end{equation}
by
\begin{equation}
\chi^{(3)}(\boldsymbol{k}+\boldsymbol{k}^{\ }_{AF},k^{\ }_z+\pi,\omega;\beta)=
\frac{S^2g}{Nc}
G(\boldsymbol{k},k^{\ }_z,\omega).
\label{eq: spin susceptibility}
\end{equation}
In the saddle-point equation, 
which becomes exact in the limit $N\rightarrow\infty$,
the ansatz $\lambda^{\ }_{*}=\mathrm{const}$ 
is chosen in order to minimize the action 
$S^{\ }_{\mathrm{eff}}[\lambda]$.
This leads to the saddle-point equation 
\begin{subequations}
\label{eq: saddle-point equation}
\begin{equation}
\begin{split}
1 = 
\frac{g}{c}
\widehat{G}^{(0)}_{j,j}(x,x;\lambda^{\ }_{*}),
\label{eq: saddle-point equation a}
\end{split}
\end{equation}
where\cite{footnote: Fourier conv} 
\begin{eqnarray}
\widehat{G}^{(0)}_{j,j}(x,x;\lambda^{\ }_{*})&=&
\frac{1}{\beta}
\sum_{\varpi=2\pi n/\beta}
\int_{\boldsymbol{p},p^{\ }_{z}}\!\!\!\!
\widehat{G}^{(0)}(\boldsymbol{p},p^{\ }_z,\varpi;\lambda^{\ }_{*})
\nonumber\\
&&
\label{eq: integral form of G0}
\end{eqnarray}
with
\begin{eqnarray}
&&
\widehat{G}^{(0)}(\boldsymbol{p},p^{\ }_z,\varpi;\lambda^{\ }_{*}):=
\frac{
1  
     }
     {
\frac{\varpi^2}{c^2}
+
\boldsymbol{p}^2
+
2\alpha(1-\cos p^{\ }_z)
+
\lambda^{\ }_*
     }.
\nonumber\\
&&
\label{eq: integral form of G0 c}
\end{eqnarray}
\end{subequations}
The saddle-point equation~(\ref{eq: saddle-point equation}) 
is equivalent to the constraint 
$\left\langle \boldsymbol{\sigma}^2_{j}(x)\right\rangle =1$. 
It can either be solved for any temperature that admits 
a nonvanishing and positive solution
$\lambda^{\ }_{*}>0$, 
which is then related to the $N=\infty$ approximation
of the temperature-dependent correlation length 
\begin{eqnarray}
\xi\equiv
\xi^{(3)}_{\pi,\pi,\pi}=
\sqrt{\frac{1}{\lambda^{\ }_{*}}},
\end{eqnarray}
or with $\lambda^{\ }_*=0$ 
in order to determine the $N=\infty$ approximation of the
critical temperature $T^{\ }_{AF}=1/\beta^{\ }_{AF}$.

We will use the saddle-point equation~(\ref{eq: saddle-point equation}) 
to verify the existence of the three universal functions defined 
in Sec.~\ref{subsec: scaling functions} 
in the $N=\infty$ approximation.

\section{
N\'eel temperature in the \texorpdfstring{$N=\infty$}{N=Infinity} limit
           }
\label{sec: Neel T in the N=infty limit}

The saddle point~(\ref{eq: saddle-point equation}) 
is UV divergent. We follow Ref.~\onlinecite{Chubukov94} and use
the relativistic Pauli-Villars regularization with cutoff $\Lambda\gg c\beta$
for the propagator in Eq.~(\ref{eq: saddle-point equation}).
The form of the saddle point~(\ref{eq: saddle-point equation}) 
then depends on the ratio between the coupling $g$ 
 and its critical value in the two-dimensional QNLSM
\begin{eqnarray}
g^{\ }_{c}:=\frac{4\pi}{\Lambda}.
\end{eqnarray}

\begin{figure}
\epsfig{file=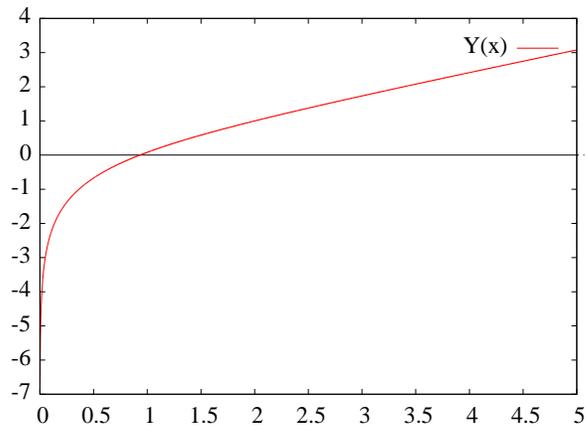,width=0.45\textwidth}
\caption{\label{fig: Y(x)}
(Color online)
Plot of the auxiliary function $Y(x)$ defined in Eq.~(\ref{eq: definition of Y})
as obtained by numerical integration.
        }
\end{figure}

When $1/g-1/g_c^{\ }>0$,
the saddle point~(\ref{eq: saddle-point equation})  
for the N\'eel temperature reduces to
\begin{subequations}
\label{eq: N=oo equation for Neel temperature A}
\begin{equation}
\frac{
2\pi\widetilde{\rho}^{\ }_{s}\beta^{\ }_{AF}
     }
     {
N
     }= 
-
Y\big(c\beta^{\ }_{AF}/\xi'\big),
\end{equation}
where we have introduced the renormalized spin stiffness
\begin{eqnarray}
\widetilde{\rho}^{\ }_{s}:=
cN
\left(
\frac{1}{g}
-
\frac{1}{g^{\ }_{c}}
\right).
\label{eq: def renormalized spin stiffness}
\end{eqnarray}
\end{subequations}
If $1/g_c^{\ }-1/g>0$,
the saddle-point~(\ref{eq: saddle-point equation})  
for the N\'eel temperature reads
\begin{subequations}
\label{eq: N=oo equation for Neel temperature B}
\begin{equation}
\Delta\ \beta^{\ }_{AF}=
+2Y\big(c\beta^{\ }_{AF}/\xi'\big),
\end{equation}
with the quantum disordered spin gap
\begin{eqnarray}
\Delta:=
4\pi c
\left(
\frac{1}{g^{\ }_{c}}
-
\frac{1}{g}
\right)
\label{eq: def renormalized spin gap}
\end{eqnarray}
\end{subequations}
in the two-dimensional QNLSM.
In order to get
Eq.~(\ref{eq: N=oo equation for Neel temperature A}) 
and Eq.~(\ref{eq: N=oo equation for Neel temperature B}), 
we first performed the summation over Matsubara frequencies
followed by the momentum integration using the Pauli-Villars regularization scheme.
In doing so we find the function
\begin{subequations}
\begin{equation}
Y(x):=
\int\limits_0^\pi 
\frac{d\varphi}{\pi}\, 
\ln\left[2\sinh\big(x\sin\varphi\big)\right]
\label{eq: definition of Y}
\end{equation}
which is monotonically increasing for $x>0$, 
has a first-order zero at $x^{\ }_0\simeq 0.93$,
and the asymptotes
\begin{equation}
Y(x)
\simeq
\left\{
\begin{array}{ll}
\ln x
+ 
\frac{x^2}{12},
& 
x\ll 1,
\\
&\\
y^{\ }_1 
(x-x^{\ }_0), 
& 
x\simeq x^{\ }_0,
\\
&\\
\frac{2x}{\pi} 
-
\frac{\gamma}{x}, 
& x\gg 1,
\end{array}
\right.
\label{eq: assymptotics of Y}
\end{equation}
\end{subequations}
with the derivative
$y^{\ }_{1}\equiv Y'(x^{\ }_0)\simeq 1.22$ 
and the constant
$\gamma\simeq 0.53$. 
We plot in Fig.~\ref{fig: Y(x)}
the function $Y(x)$.
As we shall show in the sequel,
these three asymptotic regimes
allow us to identify the three regimes
(\ref{eq: regime i}),
(\ref{eq: regime ii}),
and
(\ref{eq: regime iii})
for which the N\'eel temperature is finite.

\begin{figure}
\epsfig{file=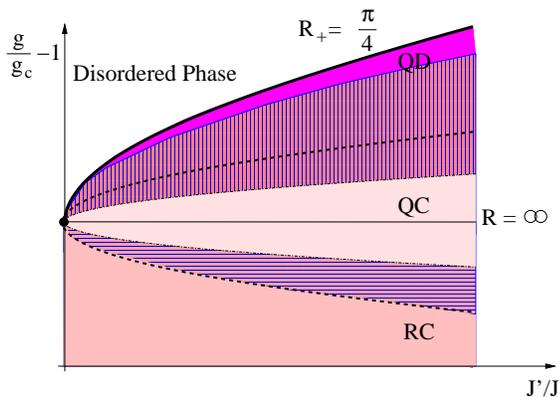,width=0.45\textwidth}
\caption{
\label{fig: Fig illustrating J'/J to 0 limit}
(Color online)
Projection of the phase diagram of the 
quasi-two-dimensional QNLSM into 
the plane spanned by 
$g/g_c-1$ and $J'/J$.
The upper bold line is the phase boundary between
the AF ordered and the paramagnetic phases
corresponding to the fixed ratio $R_+ = \pi/4$.
The hatched regions denote the crossover between the QD and QC regimes,
on the one hand, and between the QC and RC regimes, on the other hand.
Each curve represents a possible trajectory
with fixed $R_{\pm}$ used in order to construct the scaling functions
in the limit $J'/J\to 0$; the temperature along the 
curves is always fixed to $T_{AF}$.
}
\end{figure}

The limit $J'/J\to0$ always probes
the RC regime (\ref{eq: regime i}) provided the underlying
two-dimensional subsystem is not quantum disordered, i.e.,
it satisfies the condition $g<g^{\ }_{c}$.
The limit $J'/J\to0$ can only probe 
the QC regime (\ref{eq: regime ii})
or the QD regime (\ref{eq: regime iii})
after fine tuning.
This is so because, for any value $g/g^{\ }_{c}\neq1$,
the unconstrained limit $J'/J\to0$ 
brings the quasi-two-dimensional system either 
in the regime (\ref{eq: regime i}), 
that is dominated by two-dimensional classical fluctuations,
when $g/g^{\ }_{c}<1$ 
or in the quasi-two-dimensional paramagnetic phase,
that is dominated by two-dimensional quantum fluctuations at very low
temperatures, when $g/g^{\ }_{c}>1$.  
When $g/g^{\ }_{c}<1$, 
the QC regime (\ref{eq: regime ii}) can only be probed in the
$J'/J\to0$ limit if the ratio of length scales
\begin{subequations}
\label{eq: def R-+}
\begin{equation}
R^{\ }_{-}:= 
\frac{cN/(4\pi\widetilde{\rho}^{\ }_{s})}{\xi'}
\label{eq: definition of R-}
\end{equation}
is held fixed.
When  $g/g^{\ }_{c}>1$, 
the QC regime (\ref{eq: regime ii}) 
or the QD regime (\ref{eq: regime iii}) 
can only be probed in the
$J'/J\to0$ limit if the ratio of length scales
\begin{equation}
R^{\ }_{+}:= 
\frac{c/\Delta}{\xi'}
\label{eq: definition of R+}
\end{equation}
\end{subequations}
is held fixed. 
In the latter case, 
the magnitude of $R^{\ }_{+}$ distinguishes the fate of the limit $J'/J\to0$,
i.e., whether the onset of AF long-range order is characterized by
the quantum fluctuations of regime (\ref{eq: regime ii})
or the quantum fluctuations of regime  (\ref{eq: regime iii}).
When $R^{\ }_{+}$ is below the threshold value of $\pi/4$,  we shall see that
three-dimensional AF long-range order is impossible down to 
and at zero temperature 
(see Fig.~\ref{fig: Fig illustrating J'/J to 0 limit}).

\subsection{
RC regime
           }
\label{subsec: g2c bigger g0 I}

When $g<g^{\ }_{c}$ and $c\beta^{\ }_{AF}\ll\xi'$, 
Eq.~(\ref{eq: N=oo equation for Neel temperature A}) simplifies to
\begin{equation}
\frac{c\beta^{\ }_{AF}}{\xi'}\simeq 
\exp
\left(
-
\frac{
2\pi\widetilde{\rho}^{\ }_{s}\beta^{\ }_{AF}
     }
     {
N
     }
\right).
\label{eq: Neel temperature RC-regime}
\end{equation}
Observe that, for any finite $\widetilde{\rho}^{\ }_{s}$, 
the condition~(\ref{eq: regime i}) is always met
for $J'$ small enough, as, in the limit $J'/J\rightarrow 0$, 
Eq.~(\ref{eq: Neel temperature RC-regime})
has the solution
\begin{equation}
\beta^{\ }_{AF}(J'/J )
\simeq
\frac{
N
     }
     {
2\pi\widetilde{\rho}^{\ }_{s}
     }
\left[
\frac{
1
     }
     {
2
     }     
\ln
\left(
\frac{
J
     }
     {
J'
     }
\right)
+
\ln
\left(
\frac{
4\pi\widetilde{\rho}^{\ }_{s}/N
     }
     {
c\ln(J/J')
     }
\right)
\right]
\label{eq: explicit Neel temperature RC-regime}
\end{equation}
with corrections of order $\frac{\ln\ln (J/J')}{\ln (J/J')}$.
The dependence of the N\'eel temperature on $J'/J$
shows an essential singularity at  $J'/J=0$;
expansion~(\ref{eq: explicit Neel temperature RC-regime}) is very poor.

\subsection{
QC regime
           }
\label{subsec: g2c bigger g0 II}

When $g<g^{\ }_{c}$ and $c\beta^{\ }_{AF}/\xi'\lesssim x^{\ }_{0}$, 
Eq.~(\ref{eq: N=oo equation for Neel temperature A}) becomes
\begin{eqnarray}
\frac{c\beta^{\ }_{AF}}{\xi'}\simeq 
\frac{x^{\ }_{0}}{1+1/(2y^{\ }_{1}R^{\ }_{-})},
\qquad
R^{\ }_{-}=
\frac{cN}{4\pi\widetilde{\rho}^{\ }_{c}\xi'}.
\label{eq: Neel temperature QC II}
\end{eqnarray}
Equation (\ref{eq: Neel temperature QC II}) 
is self-consistent with the assumption
$c\beta^{\ }_{AF}/\xi'\lesssim x^{\ }_{0}$
provided $R^{\ }_{-}\gg1$. 
The N\'eel temperature can be expanded in a power series
in $1/R^{\ }_{-}$ 
when approaching the 
two-dimensional quantum critical point $\widetilde{\rho}^{\ }_{s}=0$
from the two-dimensional quantum ordered side.

If
$g>g^{\ }_{c}$
and 
$c\beta^{\ }_{AF}/\xi'\gtrsim x^{\ }_{0}$,
Eq.~(\ref{eq: N=oo equation for Neel temperature B})
can be rewritten as
\begin{equation}
\frac{c\beta^{\ }_{AF}}{\xi'}\simeq 
\frac{x^{\ }_{0}}{1-1/(2y^{\ }_{1}R^{\ }_{+})},
\qquad
R^{\ }_{+}=\frac{c}{\Delta\xi'}.
\label{eq: Neel temperature QD II}
\end{equation}
In order for Eq.~(\ref{eq: Neel temperature QD II}) 
to be consistent with the assumption
$c\beta^{\ }_{AF}/\xi'\gtrsim x^{\ }_{0}$,
we have to require $R^{\ }_{+}\gg1$. 
Again, the N\'eel temperature can be expanded in a power series
in $1/R^{\ }_{+}$ when approaching 
the two-dimensional quantum critical point $\Delta=0$
from the two-dimensional quantum disordered side.

\subsection{
QD regime
           }
\label{subsec: g0 bigger g2c I}

When
$g>g^{\ }_{c}$
and 
$c\beta^{\ }_{AF}\gg\xi'$,
Eq.~(\ref{eq: N=oo equation for Neel temperature B})
simplifies to
\begin{equation}
\frac{c\beta^{\ }_{AF}}{\xi'}
\simeq
\sqrt{
\frac{
2\gamma
     }
     {
\frac{4}{\pi}
-
\frac{1}{R^{\ }_{+}} 
     }
     },
\qquad
R^{\ }_{+}=\frac{c}{\Delta\xi'},
\label{eq: Neel temperature QCII}
\end{equation}
provided $R^{\ }_{+}>\pi/4$. 
When $R^{\ }_{+}\leq\pi/4$, 
the N\'eel temperature is predicted
to vanish in the $N=\infty$ approximation.
The minimal interlayer coupling $J'_c$
for which the three-dimensional N\'eel state exists
can be obtained by solving Eq.~(\ref{eq: Neel temperature QCII})
at $\beta^{\ }_{AF}=\infty$. This gives
\begin{equation}
\frac{J'_c}{J}=
\left(\frac{\pi}{4}\frac{\Delta}{c}\right)^2.
\label{eq: definition of J'_c}
\end{equation}

\subsection{
Critical coupling
           }
\label{subsec: Critical coupling}

\begin{figure}
\epsfig{file=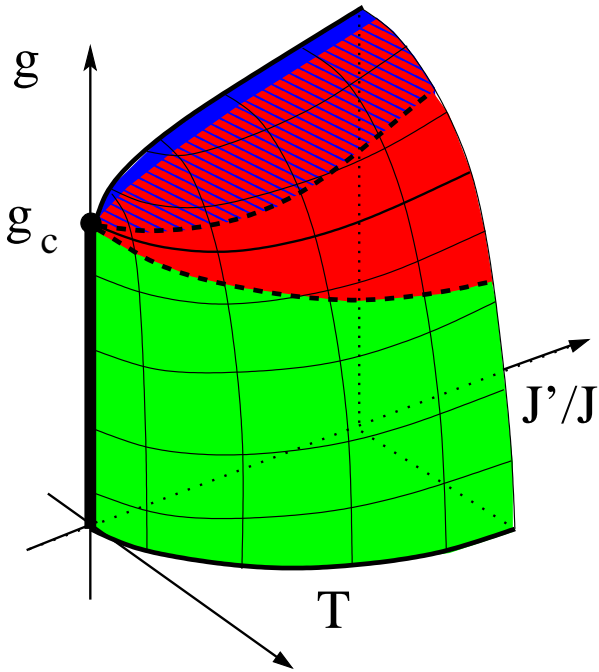,width=0.45\textwidth}
\caption{\label{fig: boundaries between the ordered and para phases}
(Color online) 
Boundaries between the AF ordered and the paramagnetic phase as a function
of the spatial anisotropy. The dependence on $J'/J$ of the
two boundaries (thick lines), one emanating from the origin $(g,T)=(0,0)$ into the
plane $g=0$, the other emanating from
the two-dimensional quantum critical point $(g,T)=(g^{\ }_{c},0)$
into the plane $T=0$,
are given by 
Eqs.~(\ref{eq: explicit Neel temperature RC-regime}) 
and (\ref{eq: critical quasi-two-dimensional coupling}), 
respectively.
The curves on the critical surface (dashed lines) lead to the
QC (lower line) and QD (upper line) behaviors.
        }
\end{figure}

The critical coupling $g^{\ }_{c}$ 
defined in Eq.~(\ref{eq: def renormalized spin stiffness})
separates the N\'eel ordered phase from the paramagnetic phase
of the two-dimensional system at zero temperature. 
We have seen that the quasi-two-dimensional ordered phase can exist
in the limit $J'/J\to0$ even though $g>g^{\ }_{c}$,
provided the two-dimensional quantum gap $\Delta$ is scaled
accordingly. This implies that any finite coupling $J'/J$ 
between planes renormalizes
$g^{ }_c$ to a larger value $\widetilde{g}^{\ }_{c}$.
The $N=\infty$ approximation for $\widetilde{g}^{\ }_{c}$ 
can be obtained from Eq.~(\ref{eq: Neel temperature QCII})
after inserting in $c/\xi'=\pi\Delta/4$
the renormalized gap
$\Delta$ from Eq.~(\ref{eq: def renormalized spin gap}).
One finds the dependence on $J'/J$,
\begin{equation}
\frac{1}{\widetilde{g}^{\ }_{c}}\simeq 
\frac{
1
     }
     {
g^{\ }_{c}
     }
- 
\frac{
1 
     }
     {
\pi^2
     }
\sqrt{\frac{J'}{J}}
+
\cdots,
\label{eq: critical quasi-two-dimensional coupling}
\end{equation}
that is characterized by an essential singularity at $J'/J=0$.
The boundaries
(\ref{eq: explicit Neel temperature RC-regime})
and
(\ref{eq: critical quasi-two-dimensional coupling})
are depicted in Fig.~\ref{fig: boundaries between the ordered and para phases}.

\section{
Scaling laws in the \texorpdfstring{$N=\infty$}{N=Infinity} limit
        }
\label{sec: Scaling laws in the N=infty limit}

We are now going to verify the claims
(\ref{eq: claim for F1}),
(\ref{eq: claim for F2}),
and
(\ref{eq: claim for F3})
in the $N=\infty$ approximation.

\subsection{
First scaling law
           }
\label{subsec: First scaling law}

One way to proceed in the case of Eq.~(\ref{eq: claim for F1}) is as follows.
We solve the saddle-point equation
(\ref{eq: saddle-point equation})
a first time with $\lambda^{\ }_{*}=0$ but $J'/J>0$ to determine 
the N\'eel temperature, 
as was done in Sec.\ \ref{sec: Neel T in the N=infty limit}.
We then solve the saddle-point equation
(\ref{eq: saddle-point equation})
a second time but now with $\lambda^{\ }_{*}=J'/J=0$ 
to extract the two-dimensional correlation length $\xi^{(2)}(\beta^{\ }_{AF})$
at the N\'eel temperature, which in turn determines 
$\chi^{(2)}_{s}(\beta^{\ }_{AF})$.  
Alternatively, we can extract $\xi^{(2)}(\beta^{\ }_{AF})$ from
\begin{equation}
\widehat{G}^{(0)}_{j,j}\Big(x,x;\lambda=0\Big)= 
\left.
\widehat{G}^{(0)}_{j,j}\left(x,x;\left(\xi^{(2)}\right)^{-2}\right)
\right|^{\ }_{\alpha=0},
\label{eq: G = G}
\end{equation}
where a matrix element of the $N=\infty$ approximation to the propagator 
$\widehat{G}^{(0)}$ 
is explicitly given in Eq.~(\ref{eq: integral form of G0 c}).
This equation has the advantage of being cutoff independent in the 
Pauli-Villars regularization scheme.

After summing over the frequencies and
integrating over the two-dimensional momenta,
Eq.~(\ref{eq: G = G}) reduces to
\begin{equation}
\frac{\xi'}{\xi^{(2)}(\beta^{\ }_{AF})}=
2
\left(
\frac{
\xi'
     }
     {
c\beta^{\ }_{AF}
     }
\right)
\arcsinh
\left(
\frac{1}{2}
e^{Y\big(c\beta^{\ }_{AF}/\xi'\big)}
\right).
\label{eq: ration xi'/xi^{(2)}}
\end{equation}
We insert this ratio into
\begin{equation}
J'\chi^{(2)}_{s}(\beta^{\ }_{AF})=
J'\times
\frac{S^2 g}{Nc}
\left(
\xi^{(2)}(\beta^{\ }_{AF})
\right)^{2}
\end{equation}
where we made use of Eq.~(\ref{eq: spin susceptibility})
using the $N=\infty$ approximation of the
Green function with $\alpha=0$.
After trading $J'$ for $J$ and $\xi'$ defined in 
Eq.~(\ref{eq: definition of xi prime}), 
we find 
\begin{eqnarray}
J'\chi^{(2)}_{s}(\beta^{\ }_{AF})&=&
\frac{S^2 g J}{Nc}
\left(
\frac{\xi^{(2)}(\beta^{\ }_{AF})}{\xi'}
\right)^2
\label{eq: first scaling law, wrong variable a}
\\
&=&
\frac{\big(c\beta^{\ }_{AF}/\xi'\big)^2}
{
4\arcsinh^2
\left( e^{Y\big(c\beta^{\ }_{AF}/\xi' \big)}/2\right)
}.
\label{eq: first scaling law, wrong variable b}
\end{eqnarray}
Equation~(\ref{eq: first scaling law, wrong variable b})
tells us that the function $J'\chi^{(2)}_{s}$
is a scaling function of the scaling variable
$c\beta^{\ }_{AF}/\xi'$.
This is not $F^{\ }_{1}$ from Eq.~(\ref{eq: claim for F1}), 
since we are looking for a function 
of $c\beta^{\ }_{AF}/\xi^{(2)}(\beta^{\ }_{AF})$
and its limiting value when $J'/J\to0$ has yet to be taken.

\subsubsection{
Renormalized classical regime
              }
\label{subsec: The renormalized classical regime}

In the RC regime~(\ref{eq: regime i}), 
where $c\beta^{\ }_{AF}/\xi'\ll 1$, and after expanding
the right-hand side of Eq.~(\ref{eq: ration xi'/xi^{(2)}}) 
in powers of $c\beta^{\ }_{AF}/\xi'$, we get
\begin{equation}
\frac{\xi'}{\xi^{(2)}(\beta^{\ }_{AF})}
\simeq
1
+
\frac{
1
     }
     {
24
     }
\left(
\frac{c\beta^{\ }_{AF}}{\xi'}
\right)^2
+
\cdots
\label{eq: ratio xi' / xi^{2d}, RC}
\end{equation}
and from Eq.~(\ref{eq: first scaling law, wrong variable b})
\begin{equation}
J' \chi^{(2)}_{s}(\beta^{\ }_{AF})=
1
-
\frac{1}{12}
\left(
\frac{c\beta^{\ }_{AF}}{\xi'}
\right)^2
+
\cdots.
\label{eq: intermediary power exp}
\end{equation}
According to Eq.~(\ref{eq: ratio xi' / xi^{2d}, RC}), 
we can replace $\xi'$ by $\xi^{(2)}(\beta^{\ }_{AF})$ 
in Eq.~(\ref{eq: intermediary power exp})
to the first nontrivial order,
\begin{equation}
J'\chi^{(2)}_{s}(\beta^{\ }_{AF})=
1
-
\frac{1}{12}
\left(\frac{c\beta^{\ }_{AF}}{\xi^{(2)}(\beta^{\ }_{AF})}\right)^2
+
\cdots.
\label{eq: first scaling law as function of c beta/ xi2-dimensional}
\end{equation}
We conclude that,
in the $N=\infty$ limit,
$J'\chi^{(2)}_{s}(\beta^{\ }_{AF})$ equals the scaling
function
\begin{subequations}
\begin{eqnarray}
\mathcal{F}^{(\mathrm{RC})}_1(x):= 
1 
- 
\frac{x^2}{12}
+
\cdots
\label{eq: RC scaling function 1}
\end{eqnarray}
with the scaling variable
\begin{eqnarray}
x:=
\frac{c\beta^{\ }_{AF}}{\xi^{(2)}(\beta^{\ }_{AF})}
\label{eq: RC scaling variable 1}
\end{eqnarray}
\end{subequations}
in the RC regime (\ref{eq: regime i}).
The universal scaling function $F_1$ is then obtained 
from $\mathcal{F}^{\mathrm{RC}}_1(x)$ 
taking the limit $J'/J\to0$. 
Observe that 
we do not expect the function $\mathcal{F}^{\mathrm{RC}}_1(x)$ 
to be universal in general. For example, this function is modified
by adding longer range interlayer
couplings in Eq.~(\ref{eq: S = S1}).\cite{footnote: nnn nnnn etc}

For comparison with numerical simulations, it is instructive to compute 
this relation as a function of $\alpha=J'/J$. 
Inserting the N\'eel temperature from Eq.~(\ref{eq: explicit Neel temperature RC-regime})
in Eq.~(\ref{eq: first scaling law as function of c beta/ xi2-dimensional})
leads to
\begin{equation}
J'\chi^{(2)}_{s}(\beta^{\ }_{AF})\simeq
1
-
\frac{1}{12}
\frac{
c^{2}N^{2}
     }
     {
\left
(4\pi\widetilde{\rho}^{\ }_{s}
\right)^2
     }
\frac{J'}{J}
\ln^2
\left(
\frac{J}{J'}
\right)
+\cdots.
\label{eq: first scaling law as function of J'/J}
\end{equation}
The universal value in the limit $J'/J\rightarrow 0$ is thus established, 
since 
\begin{subequations}
\begin{eqnarray}
F^{\ }_{1}(0)=1
\label{eq: universal limit in RC for F1}
\end{eqnarray}
is independent of any microscopic details, 
i.e., the spin stiffness $\widetilde{\rho}^{\ }_{s}$ 
or the spin-wave velocity $c$.
For comparison, the RPA predicts 
\begin{eqnarray}
F^{\mathrm{(RPA)}}_{1}(0)=1/2,
\end{eqnarray} 
while the MC calculations from Ref.~\onlinecite{Yasuda05} 
gives
\begin{eqnarray}
F^{\mathrm{(MC)}}_{1}(0)\simeq  
0.77.
\end{eqnarray} 
\end{subequations}

We show in Appendix~\ref{app: Quasi-one-dimensional Ising model} 
that the form of the nonuniversal corrections in 
Eq.~(\ref{eq: first scaling law as function of J'/J}) 
are similar to the corrections obtained 
in the quasi-one-dimensional Ising model.

The factor  $cN/(4\pi\widetilde{\rho}^{\ }_{s})$ 
scales as $1/S$.
As expected, the nonuniversal corrections get smaller for bigger spins.
We show in Sec.~\ref{sec: Scaling laws in the N=S=infty limit}
that the nonuniversal corrections vanish
in the $N=\infty$ approximation to the classical Heisenberg AF. 

\begin{figure}
\epsfig{file=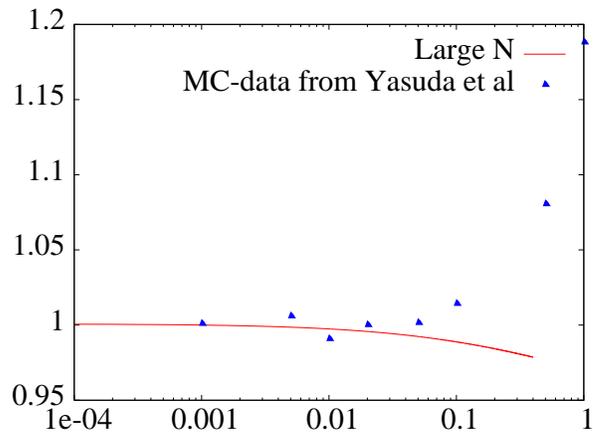,width=0.45\textwidth}
\caption{\label{fig: comparison with results of Yasuda}
(Color online)
The scaling function $F^{\ }_1$ is plotted vs $J'/J$ for spin $1/2$.
The curve represents the $N=\infty$ approximation%
~(\ref{eq: first scaling law as function of c beta/ xi2-dimensional}),
with the N\'eel temperature obtained from numerical solution of 
Eq.~(\ref{eq: Neel temperature RC-regime}).
The points are the MC data from 
Yasuda {\it et al.} in Ref.~\onlinecite{Yasuda05}. 
In both cases, the normalization is such that the curves go through 1 for 
$J'/J=10^{-3}$.
        }
\end{figure}

The result~(\ref{eq: first scaling law as function of J'/J}) 
is only reliable for very small $J'/J$,
due to the limited quality of 
the approximation~(\ref{eq: explicit Neel temperature RC-regime}). 
To compare the $N=\infty$ approximation with MC simulations, 
it is best to rely on a numerical solution of 
Eq.~(\ref{eq: Neel temperature RC-regime}) 
for the N\'eel temperature.
As can be seen from Fig.~\ref{fig: comparison with results of Yasuda},
for $J'/J<0.1$, the $N=\infty$ approximation%
~(\ref{eq: first scaling law as function of c beta/ xi2-dimensional})
of the first scaling law obeys a limiting behavior as $J'/J\rightarrow 0$
similar to the MC results from Ref.~\onlinecite{Yasuda05}.
For $J'/J>0.1$, the magnitude of the deviation of the 
effective coordination number from its universal asymptotic value as $J'/J\to0$ 
can be explained using the empirical 
formula for the N\'eel temperature proposed in Ref.~\onlinecite{Yasuda05}.
However, the sign of the nonuniversal correction in 
Eq.~(\ref{eq: first scaling law as function of c beta/ xi2-dimensional})
is wrong. We do not expect the quasi-two-dimensional QNLSM used here 
to be valid for the description of the system at $J'/J>0.1$
as the effective interlayer separation $\xi'$ 
is only a few lattice spacing large. Instead,
it is likely that it is necessary to also consider the lattice structure
of the system in each plane to obtain an
appropriate description.

\subsubsection{
Quantum critical regime
              }
\label{subsec: Quantum critical regime}

The condition $\Delta=0$ fine tunes 
the underlying two-dimensional system to be at
a quantum critical point.
From Eqs.~(\ref{eq: N=oo equation for Neel temperature B}) 
and (\ref{eq: assymptotics of Y}), 
we then have $c\beta^{\ }_{AF}/\xi'=x^{\ }_0$.
Taking advantage of this relation
in Eq.~(\ref{eq: ration xi'/xi^{(2)}}) gives
\begin{subequations}
\begin{eqnarray}
\frac{\xi'}{\xi^{(2)}(\beta^{\ }_{AF})}= 
\frac{\Theta}{x^{\ }_{0}},
\label{eq: ratio xi' to xi2dim if QC}
\end{eqnarray}
where 
\begin{eqnarray}
\Theta:= 
2\ln\frac{1+\sqrt{5}}{2}.
\label{eq: def Theta}
\end{eqnarray}
\end{subequations}
The right-hand side of Eq.~(\ref{eq: first scaling law, wrong variable a})
thus becomes
\begin{eqnarray}
J'\chi^{(2)}_{s}(\beta^{\ }_{AF})&=&
\frac{x^{2}_{0}}{\Theta^2}.
\label{eq: F1 exactly at QC}
\end{eqnarray}
Chubukov {\it et al.} in Ref.~\onlinecite{Chubukov94}
have shown that
\begin{eqnarray}
\frac{
c\beta
     }
     {
\xi^{(2)}(\beta)
     }=\Theta
\end{eqnarray}
in the $N=\infty$ approximation when $\Delta=0$. 
The universal function $F^{\ }_{1}$ from Eq.~(\ref{eq: claim for F1})
thus takes the value
\begin{eqnarray}
F^{\ }_{1}(\Theta)=
\frac{x^{2}_{0}}{\Theta^2}.
\end{eqnarray}
exactly at the QC point $\Delta=0$.

In the scaling limit $J'/J\to0$ with the ratios
$R^{\ }_{\mp}$ from 
Eqs.~(\ref{eq: definition of R-}) 
and (\ref{eq: definition of R+}) held fixed,
Eqs.~(\ref{eq: ratio xi' to xi2dim if QC})
and (\ref{eq: F1 exactly at QC})
become
\begin{equation}
\frac{\xi'}{\xi^{(2)}(\beta^{\ }_{AF})}= 
\frac{\Theta}{x^{\ }_{0}}
+
\frac{\kappa}{R^{\ }_{\mp}}
+
\mathcal{O}\big(R^{-2}_{\mp}\big)
\end{equation}
and
\begin{equation}
J'\chi^{(2)}_{s}(\beta^{\ }_{AF})=
\frac{x^{2}_{0}}{\Theta^2}
-
\frac{\kappa}{R^{\ }_{\mp}}
+
\mathcal{O}\big(R^{-2}_{\mp}\big),
\end{equation}
respectively,
where
\begin{eqnarray}
\kappa:= 
\frac{
\Theta
+
2y^{\ }_{1}/\sqrt{5}
     }
     {
x^{\ }_{0}
     }.
\end{eqnarray}

\subsubsection{
Quantum disordered regime
              }
\label{subsec: Quantum disordered regime}

We close with the QD regime~(\ref{eq: regime iii})
for which $c\beta^{\ }_{AF}/\xi'\gg 1$. 
The underlying two-dimensional system is in its quantum disordered phase 
with 
\begin{eqnarray}
\xi^{(2)}(T=0)=c/\Delta.
\end{eqnarray}
Taking the scaling limit $J'/J\to0$ with 
the ratio $R^{\ }_{+}$ from Eq.~(\ref{eq: definition of R+}),
\begin{eqnarray}
R^{\ }_{+}&=&
\frac{c}{\Delta\xi'}
\nonumber\\
&=&
\xi^{(2)}(T=0)/\xi'
\nonumber\\
&=&
\xi^{(2)}(\beta^{\ }_{AF})/\xi'
+
\mathcal{O}\left(e^{-c\beta^{\ }_{AF}/\Delta}\right),
\end{eqnarray}
held fixed turns Eq.~(\ref{eq: first scaling law, wrong variable a}) 
into
\begin{equation}
J'\chi^{(2)}(\beta^{\ }_{AF})\simeq 
R^{2}_{+}
\end{equation}
up to exponentially small corrections in $(R^{\ }_{+}-\pi/4)^{-1/2}$.

\subsubsection{
Summary
              }
\label{subsec: Summary}

To sum up, we have the $N=\infty$ approximation 
\begin{equation}
\lim_{J'/J\to0}
J'\chi^{(2)}(\beta^{\ }_{AF})
\simeq
\left\{
\begin{array}{ll}
1,
& 
\mbox{ RC regime},
\\&\\
\frac{
x^{2}_{0}
     }
     {
\Theta^{2}
     }
-
\frac{\kappa}{R^{\ }_{\mp}},
& 
\mbox{ QC regime},
\\&\\
R^{2}_{+},
& 
\mbox{ QD regime},
\end{array}
\right.
\end{equation}
to the first universal function~(\ref{eq: claim for F1}),
where $R^{\ }_{\mp}\gg 1$ in the QC regime 
while $R^{\ }_{+}\gtrsim\pi/4$ in the QD regime. 
The result 
$F^{\ }_{1}(0)=1$ is exact in the RG regime~(\ref{eq: claim for F1}).
The result 
$
F^{\ }_{1}\big(c\beta^{\ }_{AF}/x^{(2)}(\beta^{\ }_{AF})\big)=
\frac{
x^{2}_{0}
     }
     {
\Theta^{2}
     }
-
\frac{\kappa}{R^{\ }_{\mp}}
$
is accurate up to corrections of order $R^{-2}_{\mp}$
in the QC regime~(\ref{eq: regime i}).
The result 
$
F^{\ }_{1}\big(c\beta^{\ }_{AF}/x^{(2)}(\beta^{\ }_{AF})\big)=
R^{2}_{+}
$
is accurate up to exponentially small corrections in $(R^{\ }_{+}-\pi/4)^{-1/2}$
in the QD regime~(\ref{eq: regime ii}).
The RC regime~(\ref{eq: regime iii})
is the only one for which we were able
to compute the nonuniversal corrections
to the limit $J'/J\to0$. They are of second order in the variable
$c\beta^{\ }_{AF}/\xi^{(2)}(\beta^{\ }_{AF})$
but nonanalytic in the variable $J'/J$, namely of order
$(J'/J)\ln^{2}(J/J')$.

\subsection{
Second scaling law
           }
\label{subsec: Second scaling law}

The second universal scaling law~(\ref{eq: claim for F2}), 
which involves the three-dimensional spin susceptibility,
is established as follows. 
In the $N=\infty$ approximation,
\begin{equation}
\chi^{(3)}\big(\pi,\pi,0,\omega=0;\beta^{\ }_{AF}\big)=
\frac{S^{2}g}{cN}\frac{1}{4\alpha},
\end{equation}
which immediately leads to the second universal relation
\begin{equation}
J'\chi^{(3)}_{s}(\beta^{\ }_{AF})= 
\frac{1}{4},
\end{equation}
in the RC, QC, and QD regimes.

\subsection{
Third scaling law
           }
\label{subsec: Third scaling law}

We close by establishing the third universal scaling law~(\ref{eq: claim for F3}),
which involves the instantaneous structure factor. 
We insert the definition~(\ref{eq: spin susceptibility}) in
Eq.~(\ref{eq: definition of S}), 
and use the $N=\infty$ approximation to the correlation function.
After performing the frequency integration, we find
\begin{equation}
S(\boldsymbol{k}^{\ }_{AF},k^{\ }_{z}=0,t=0;\beta)=
\frac{S^{2}g}{4N}\xi'\coth\big(c\beta/\xi'\big).
\end{equation}
The last universal relation is then established,
\begin{subequations}
\begin{equation}
\beta^{\ }_{AF}J'
S(\boldsymbol{k}^{\ }_{AF},k^{\ }_{z}=0,t=0;\beta^{\ }_{AF})=
\mathcal{F}^{\ }_3\big(c\beta^{\ }_{AF}/\xi'\big),
\label{eq: third universal relation N=oo}
\end{equation}
where in the $N=\infty$ approximation,
\begin{equation}
\mathcal{F}^{\infty}_3(x):=
\frac{x}{4}\coth x
\end{equation}
and
\begin{eqnarray}
x:=
\frac{c\beta^{\ }_{AF}}{\xi'}.
\end{eqnarray}
\end{subequations}

\noindent
The universal scaling function $F_3$ is then obtained 
from $\mathcal{F}^{\infty}_3(x)$ taking the limit $J'/J\to0$. 
As for the first scaling law, we do not expect the function $\mathcal{F}^{\infty}_3(x)$
 to be universal in general.
The behavior of $\mathcal{F}^{\infty}_{3}(x)$ when 
$x\ll 1$,
$x\sim1$,
and
$x\gg1$
distinguishes the RC, QC, and QD regimes.
Observe in this context that, as
$\mathcal{F}^{\infty}_3(x)\rightarrow 1/4+\mathcal{O}(x^2)$ %
in the limit of small $x$,
the function $F^{\ }_3$ 
is the same as $F^{\ }_{2}$ in the RC regime,
as expected from the quasielastic approximation~\cite{Birgeneau71,Keimer92}
\begin{equation}
\chi(\boldsymbol{q},\omega=0;\beta)
\simeq 
-\beta S(\boldsymbol{q},t=0;\beta).
\end{equation} 
In the QD regime, the scaling variable $x$ can
become arbitrary large as $R_{+}\to \pi/4$. 
[In particular, this is the case when $J'\to J'_c$, 
see Eq.~(\ref{eq: definition of J'_c}).]
Consequently, the function $\mathcal{F}^{\ }_3(x)\simeq x/4$ on the %
right hand side of Eq.~(\ref{eq: third universal relation N=oo})
diverges when the ratio $R^{\ }_{+}$ 
that defines how the limit $J'\to 0$ is taken
approaches $\pi/4$.

\section{
Scaling laws in the \texorpdfstring{$N=S=\infty$}{N=S=Infinity} limit
        }
\label{sec: Scaling laws in the N=S=infty limit}

The three components of the spin operators 
with spin quantum number $S$ commute in the limit $S\to\infty$. 
The results of Yasuda \textit{et al.}\
suggest that the effective coordination number\
(\ref{eq: def zeta}) is independent of $S$ in the limit $J'/J\to0$.
This limiting value of the effective coordination number is thus the same
irrespective of $S$ taking values deep in the quantum regime, $S=1/2$, 
or classical regime, $S=\infty$. 
If so, it is instructive to compare our analysis of the quasi-two-dimensional QNLSM%
~(\ref{eq: S = S0+S1})
with that of the  quasi-two-dimensional NLSM.
The action of the quasi-two-dimensional NLSM
is obtained from Eq.~(\ref{eq: S = S0+S1})
by dropping all references to the imaginary time,
setting $S=1$ and $c=1$,
and replacing the integral over $\tau$ by $1/T$,
\begin{equation}
S[\boldsymbol{\sigma}]:=
\frac{\rho^{\ }_{s}}{2T}\sum_j \int d^{2}\boldsymbol{x}\, 
\left[
\big(\nabla\boldsymbol{\sigma}_j\big)^2
+\frac{J'}{J}\boldsymbol{\sigma}_j\cdot\boldsymbol{\sigma}_{j+1}
\right]
\end{equation}
with $\rho^{\ }_{s}\equiv J$.
Again, the constraint $\boldsymbol{\sigma}^{2}_{j}(\boldsymbol{x})=1$ 
can be removed in favor of the Lagrange multiplier $\lambda$.
If the original fields $\boldsymbol{\sigma}$ are integrated out,
one obtains the effective action 
\begin{subequations}
\begin{equation}
S^{\ }_{\mathrm{eff}}[\lambda]=
-
\frac{1}{2gT}
\sum_{j}
\int d^2 x \, \lambda^{\ }_{j}
+
\frac{1}{2}
\operatorname{Tr} 
\ln\big(-\boldsymbol{\nabla}^2-\alpha\Delta^{\ }_z+\lambda^{\ }_{j}\big)
\end{equation}
where
\begin{eqnarray}
g=N/\rho^{\ }_{s}
\end{eqnarray}
\end{subequations}
that should be compared with Eqs.~(\ref{eq: effective action}) and (\ref{eq: def g}).
The spin susceptibility is now related to the two-point correlation functions by
\begin{equation}
\chi^{(3)}(\boldsymbol{k}+\boldsymbol{k}_{AF},\pi+k_z;\beta)=
\frac{g}{N}
G(\boldsymbol{k},k_z;\beta)
\label{eq: classical spin susceptibility}
\end{equation}
that should be compared with Eq.~(\ref{eq: spin susceptibility}).

To compute the first scaling law in the $N=\infty$ approximation, we proceed 
as in the quantum case.
We thus solve directly the classical counterpart to Eq.~(\ref{eq: G = G}) 
for the ratio $\xi'/\xi^{(2)}$ 
with $\xi^{(2)}$ evaluated at the N\'eel temperature. We find
\begin{equation}
\frac{\xi'}{\xi^{(2)}(\beta^{\ }_{AF})}=1.
\end{equation}
Hence the first scaling law
in the $N=\infty$ approximation simply reads
\begin{subequations}
\begin{equation}
J'\chi^{(2)}_{s}(\beta^{\ }_{AF})=
\mathcal{F}^{\ }_{1}(x)
\end{equation}
with
\begin{equation}
\mathcal{F}^{\ }_{1}(x)=1
\end{equation}
and
\begin{equation}
x=\frac{J\beta^{\ }_{AF} a}{\xi^{(2)}(\beta^{\ }_{AF})}.
\end{equation}
\end{subequations}
The counterpart to the universal function~(\ref{eq: claim for F1})
obtained in the limit $J'/J\rightarrow 0$ is then
\begin{eqnarray}
F^{\ }_{1}(x)=1.
\end{eqnarray}
It agrees with the RC limit~(\ref{eq: universal limit in RC for F1}).
This is consistent with the MC study from Yasuda {\it et al.} 
in Ref.~\onlinecite{Yasuda05}. Contrary to the quantum case,
nonuniversal corrections are absent in the saddle-point approximation.
This result is also consistent with the observation
from Ref.~\onlinecite{Yasuda05} that the finite $J'/J$
nonuniversal corrections decrease with increasing $S$.

We turn to the second scaling law. 
Inserting the $N=\infty$ approximation of the two-point
correlation function in Eq.~(\ref{eq: classical spin susceptibility}) 
at the wave vector $(\pi,\pi,0)$ leads to 
\begin{equation}
J'\chi_{s}^{(3)}(\beta^{\ }_{AF})=
\frac{1}{4},
\end{equation}
as in the quantum case. 

There is no independent third scaling law in classical thermodynamic 
equilibrium since all observables are time independent by assumption.

\section{
Comparison of the scaling laws with experiments
        }

Many quasi-two-dimensional quantum Heisenberg AF are now available. 
Neutron studies have been performed on compounds with spin $S=1/2$
[\lac\ in Refs.~\onlinecite{Birgeneau99} and \onlinecite{Toader05}, 
 \src\ in Refs.~\onlinecite{Borsa92} and \onlinecite{Greven95}, 
and copper formate tetradeuterate (CFTD) in Ref.~\onlinecite{Ronnow99}]
spin $S=1$ 
(\lan\ in Ref.~\onlinecite{Nakajima95} and \kni\ in Ref.~\onlinecite{Greven95}), 
as well as spin $S=5/2$ 
(\kfe\ in Ref.~\onlinecite{Fulton94} and \rbm\ 
in Refs.~\onlinecite{Lee98} and \onlinecite{Leheny99}).
All these materials are good realizations
of the two-dimensional QNLSM above their ordering temperature $T^{\ }_{AF}$. 
In particular, the temperature dependence of the correlation length
in these systems is well explained (without free parameter) 
by the correlation length of the two-dimensional QNLSM in the RC regime and
at the three-loop order,~\cite{Chakravarty88,Chakravarty89,Chubukov94,Hasenfratz91}
\begin{equation}
\label{CHN-HN}
\frac{\xi^{(2)}(T)}{a}
=
\frac{e}{8}\frac{c/a}{2\pi\rho_s}e^{2\pi\rho_s/T}
\left[
1
-
\frac{1}{2}\left(\frac{T}{2\pi\rho_s}\right)+\cdots
\right].
\end{equation}

All these systems have an interplane exchange coupling $J'/J$ 
which is much smaller than the spin anisotropy.
The latter is either of $XY$ type (for the $S=1/2$ systems listed above)
or Ising type (for the remaining examples chosen). 
Hence the onset of AF magnetic long-range order at $T^{\ }_{AF}$
is a classical critical point that
belongs to the $XY$ or Ising universality class.
Moreover, $T^{\ }_{AF}$ is mainly determined by the (lower)
critical temperature of the two-dimensional system,
which is finite in opposition to that for the spin-isotropic 
two-dimensional quantum Heisenberg AF.
For this reason, we do not believe that the scaling laws derived 
above for a spin-isotropic quantum Heisenberg AF
can be observed in the materials mentioned above.

A family of organic quasi-two-dimensional quantum AF
has been synthesized with the
general chemical formula A$_2$CuX$_4$, where A=5CAP
(5CAP stands for 2-amino-5-chloropyridinium)
or A=5MAP 
(5MAP stands for 2-amino-5-methylpyridinium),
and X=Br or Cl.~\cite{Woodward02}
The AF Heisenberg exchange coupling $J$ in these materials
is between 6 and $10K$, and they have a N\'eel temperature of about 
$T_{AF}\simeq 0.6 J$. According to Ref.~\onlinecite{Woodward02}, 
the dominant subleading term in the Hamiltonian describing 
their magnetic response is the interlayer Heisenberg exchange coupling $J'$.
The rationale for $J'$ being the leading subdominant term
to the planar $J$ is the following mean-field argument.~\cite{Woodward02}
Consider the compound \lac\ as an example of a layered structure
in which adjacent layers are staggered,
i.e., there exists a relative in-plane displacement $(1/2,1/2)$ 
between any two neighboring layers.
Each in-plane ion has thus four equidistant neighbors in the layer
directly above it. Assume that, in any given layer,
the spin degrees of freedom occupy the sites of a square lattice
and that they are frozen in a classical N\'eel configuration.
Any in-plane spin has then four nearest-neighbors in the layer above it
whose net mean field vanishes. In contrast,
the stacking of planes is not staggered 
in the organic quasi-two-dimensional AF
described in Ref.~\onlinecite{Woodward02},
so that this cancellation mechanism does not work. 
Consequently, the ratio $J'/J$ is expected to be much larger
for A$_2$CuX$_4$ than for \lac.

The interlayer exchange parameter for A$_2$CuX$_4$ 
is estimated in Ref.~\onlinecite{Woodward02} to be $J'/J\simeq 0.08$
with the help of the formula
\begin{equation}
T^{\ }_{AF}=z J' S(S+1) \big[\xi^{(2)}(T^{\ }_{AF})/a\big]^2.
\label{eq: cond for Neel temp from NRJ}
\end{equation}
Equation~(\ref{eq: cond for Neel temp from NRJ})
can be found in Ref.~\onlinecite{Chakravarty88} 
while its quasi-one-dimensional version was obtained by 
Villain and Loveluck in Ref.~\onlinecite{Villain77}.
The N\'eel temperature of
the spin-1/2 quasi-two-dimensional Heisenberg model
is hereby estimated by balancing the thermal energy $T^{\ }_{AF}$
against the gain in Zeeman energy obtained by aligning
a planar spin $S(S+1)$ along a mean-field magnetic field of magnitude
$zJ'[\xi^{(2)}(T^{\ }_{AF})/a]^2$
that points parallel to the stacking direction.
Here $\xi^{(2)}(T_{AF})$ is the two-dimensional correlation length
estimated at the N\'eel temperature from formula~(\ref{CHN-HN})
and $z$ is the coordination number of layers.
Using their empirical formula
(which is consistent with the first scaling law in the limit $J'/J\to 0$),
Yasuda \textit{et al.} in Ref.~\onlinecite{Yasuda05}
claim the three times larger value $J'/J\simeq 0.24$.
This discrepancy is explained in 
Appendix~\ref{app:Two different estimates fo TAF}.
To our knowledge, there is no independent estimate for the interlayer coupling 
$J'/J$ deduced from measuring the spin-wave velocity
for A$_2$CuX$_4$.
We are also not aware of measurements 
of the correlation length above the critical temperature
for A$_2$CuX$_4$.

\subsection{Experimental discussion of the first scaling law}
\label{subsec: Experimental discussion of the first scaling law}

The two-dimensional spin susceptibility is inaccessible to a direct measurement 
at the N\'eel temperature, the temperature at which our scaling laws hold.
This is not to say that the two-dimensional spin susceptibility is inaccessible
at all temperatures. In fact, one would expect a window of temperature above
the  N\'eel temperature for which the three-dimensional spin  susceptibility
is well approximated by the two-dimensional one in a good quasi-two-dimensional
AF magnet. If so one could try to test the first scaling law by extrapolation
from high temperatures.

To this end, we could first attempt to use the measured values of
the N\'eel temperature and of the anisotropic spin-wave dispersion
together with computations of the two-dimensional spin-wave velocity $c$
and spin stiffness $\rho^{\ }_{s}$ from first principle,
to test the accuracy of the prediction~(\ref{eq: ratio xi' / xi^{2d}, RC})
for the ratio 
\begin{equation}
\Xi:=
\frac{\xi^{(2)}(T^{\ }_{AF})}{\xi'}
\label{eq: ratio Xi}
\end{equation}
between the two-dimensional correlation length
$\xi^{(2)}(T^{\ }_{AF})$
and the effective interlayer spacing%
~(\ref{eq: definition of xi prime}).
Indeed, it is reassuring to know that
there exists a good agreement between formula~(\ref{CHN-HN})
and measurements of the correlation length above $T^{\ }_{AF}$ in the RC regime. 

Needed are the numerator and denominator 
in the RC regime of the right-hand side of
Eq.~(\ref{eq: ratio Xi})
expressed as a function of $N$, $c$, $\rho^{\ }_{s}$, and $T^{\ }_{AF}$
to first order in the $1/N$ expansion.
The two-dimensional correlation length is~\cite{Chubukov94}
\begin{subequations}
\begin{equation}
\frac{\xi^{(2)}(T)}{a}=
\xi^{\ }_0 
\frac{c/a}{T}
\left(
\frac{(N-2)T}{2\pi\rho^{\ }_s}
\right)^{1/(N-2)}
\exp{\frac{2\pi\rho^{\ }_s}{(N-2)T}},
\label{eq: large N xi 2d stag}
\end{equation}
where the constant
\begin{equation}
\xi^{\ }_0 = 
\big(e/8\big)^{1/(N-2)}\Gamma\big(1+1/(N-2)\big)
\end{equation}
\end{subequations}
depends on $e=2.718...$ and the Gamma function.
The N\'eel temperature is the implicit solution to~\cite{Irkhin97}
\begin{subequations}
\begin{eqnarray}
\frac{4\pi\rho^{\ }_s}{(N-2)T^{\ }_{AF}}&=&
\ln \frac{T^{2}_{AF}}{c^2/a^2}
+
\frac{2}{N-2}\ln\frac{4\pi\rho_s}{(N-2)T^{\ }_{AF}}
\nonumber\\
&&
+
\frac{1.012}{N-2}
+
\ln\frac{1}{\mathcal{Z}^{\prime}}
+
\ln\frac{J}{J'}
\end{eqnarray}
with
\begin{eqnarray}
\mathcal{Z}^{\prime}&=&
\left( 
1 
- 
\frac{8}{3\pi^{2}N}
\ln \frac{c\Lambda N}{16\rho^{\ }_s}
\right)
\nonumber\\
&&
\times
\left(
1
+
\frac{1.069}{N}
\right)
\left(  
\frac{(N-2)T}{4\pi\rho^{\ }_s}  
\right)^{1/(N-2)}.
\end{eqnarray}
\end{subequations}
The dependence on the momentum cutoff $\Lambda$
of the two-dimensional correlation length
occurs through $\xi'$ only,
\begin{equation}
\frac{\xi^{(2)}(T^{\ }_{AF})}{a}=
\xi^{\ }_0 
\left(
2
e^{0.506}
\right)^{1/(N-2)}
\frac{\xi'}{a}.
\label{eq: ratio xi' / xi^{2d}, RC 1/N}
\end{equation}
One thus gets the universal number
\begin{equation}
\Xi=
\Gamma\big(1+1/(N-2)\big)
\left(
\frac{e^{1.506}}{4}
\right)^{1/(N-2)}
\end{equation}
as it is independent of the spin-wave velocity and the spin stiffness.
In particular,
$\Xi\approx 1.127$ for $N=3$. 
(In the $N=\infty$, we found $\Xi=1$.)

We have computed the ratio $\Xi$ for some spin $S=1/2$ 
quasi-two-dimensional AF using the values of 
$T^{\ }_{AF}$, $(J'/J)^{\ }_{\mathrm{sw}}$, $2\pi\rho^{\ }_s$, and $c$ 
listed in Table~\ref{tab: spin-1/2}.
Here, $(J'/J)^{\ }_{\mathrm{sw}}$ is the ratio of the 
interplane to the intraplanar exchange couplings
deduced from the spin-wave spectra
measured using inelastic neutrons scattering 
at temperatures well below the measured $T^{\ }_{AF}$.
The measured value $(J'/J)^{\ }_{\mathrm{sw}}$ is interpreted as
the multiplicative renormalization 
\begin{eqnarray}
\left(\frac{J'}{J}\right)^{\ }_{\mathrm{sw}}=
\left( 
1 
- 
\frac{8}{3\pi^{2}N}
\ln \frac{c\Lambda N}{16\rho^{\ }_s}
\right)
\left(\frac{J'}{J}\right)
\end{eqnarray}
that arises solely from quantum fluctuations.~\cite{Irkhin97}
In turn, this allows us to express
$\xi'$ in Eq.~(\ref{eq: definition of xi prime})
in terms of the microscopic parameters $a$, $\Lambda$, $J$, and $J'$
on the one hand, and the macroscopic parameters
$\rho^{\ }_{s}$ and $T^{\ }_{AF}$ on the other hand,
\begin{eqnarray}
\frac{\xi'}{a}&=&
\left(\frac{J'}{J}\right)^{-1/2}_{\mathrm{sw}}
\left(
1
+
\frac{1.069}{N}
\right)^{-1/2}
\nonumber\\
&&
\times
\left(  
\frac{(N-2)T^{\ }_{AF}}{4\pi\rho^{\ }_s}  
\right)^{-1/[2(N-2)]}.
\label{eq: xi prime from spin-wave J'/J}
\end{eqnarray}
The crossover length scale $\xi'$
in Table~\ref{tab: spin-1/2}
then follows from inserting $N=3$ and the values
for $T^{\ }_{AF}$ and $\rho^{\ }_{s}$ from Table~\ref{tab: spin-1/2}.
The same is done with Eq.~(\ref{eq: large N xi 2d stag})
to obtain $\xi^{(2)}(T^{\ }_{AF})$
in Table~\ref{tab: spin-1/2}.
One finds the measured values
$\Xi^{\ }_{\mathrm{meas}}\approx 0.49$ for \lac, 
$\Xi^{\ }_{\mathrm{meas}}\approx 0.02$ for \src,
and $\Xi^{\ }_{\mathrm{meas}}\approx 0.34$ for CFTD. 
The smallness of the measured $\Xi^{\ }_{\mathrm{meas}}$ 
compared to $\Xi\approx 1.127$
indicates that the actual transition to the ordered
phase takes place when the two-dimensional correlation length is smaller 
than the effective interlayer separation $\xi'$.
This discrepancy with the large-$N$ expansion can be understood as follows.
The compounds of Table~\ref{tab: spin-1/2} all have a
smaller anisotropy $(J'/J)^{\ }_{\mathrm{sw}}$ as compared to the $XY$ anisotropy
or to the Dzyaloshinsky-Moria term. 
Furthermore, the \src system has about the same critical
temperature as the other two compounds, although its anisotropy is three
orders of magnitude smaller. 
This indicates~\cite{Birgeneau99,Toader05,Greven95,Ronnow99}
that the phase transition is not triggered by a pure dimensional crossover
but by a combination of a symmetry and dimensional crossover that
effectively enhances the true two-dimensional correlation length
over that for the pure two-dimensional $O(3)$ QNLSM.
The same conclusions may be drawn for systems with higher spins.

We close this section
by evaluating the first scaling law in the RC regime to the order $1/N$. 
The two-dimensional spin susceptibility at the wave vector $(\pi,\pi)$ 
and at vanishing frequency
$\omega=0$ is
\begin{subequations}
\begin{eqnarray}
\chi^{(2)}_{s}(T^{\ }_{AF})= 
\mathcal{Z}
\frac{S^{2}g}{Nc}
\left(
\frac{
\xi^{(2)}(T^{\ }_{AF})
     }
     {
a
     }
\right)^{2},
\end{eqnarray}
where the two-dimensional multiplicative renormalization
$\mathcal{Z}$ is~\cite{Chubukov94,footnote: Z}
\begin{eqnarray}
\mathcal{Z}&=&
\left( 
1 
- 
\frac{8}{3\pi^{2}N}
\ln \frac{c\Lambda N}{16\rho^{\ }_s}
\right)
\\
&&
\times
\left(
1
+
\frac{0.188}{N}
\right)
\left(  
\frac{(N-2)T^{\ }_{AF}}{2\pi\rho^{\ }_s}  
\right)^{1/(N-2)}.
\nonumber
\end{eqnarray}
\end{subequations}
Remarkably, the momentum cutoff $\Lambda$ drops from the ratio
\begin{eqnarray}
\frac{\mathcal{Z}}{\mathcal{Z}^{\prime}}=
\frac{
1
+
\frac{0.188}{N}
     }
     {
1
+
\frac{1.069}{N}
     }
\times
2^{-1/(N-2)}.
\end{eqnarray}
For $N=3$, $\mathcal{Z}/\mathcal{Z}^{\prime}=0.391$. 
(In the $N=\infty$ limit, we find $\mathcal{Z}/\mathcal{Z}^{\prime}=1$.)
The $1/N$ correction to the first scaling law thus reads
\begin{eqnarray}
J'\chi^{(2)}_{s}(T^{\ }_{AF})=
\frac{\mathcal{Z}}{\mathcal{Z}^{\prime}}
\Xi^2.
\end{eqnarray}
For $N=3$,
$J'\chi^{(2)}_{s}(T^{\ }_{AF})=0.497$.
Observe that this is very close to the RPA prediction of $1/2$.

\begin{table}
\begin{tabular}{cccc}
\hline\hline
&  \lac
&  \src
& CFTD
\\
&
Refs.~\onlinecite{Birgeneau99} and \onlinecite{Toader05}
&
Ref.~\onlinecite{Greven95}
&
Ref.~\onlinecite{Ronnow99} 
\\
\hline
$J$ 
&  $1566 K$      & $1450 K$            & $73.2 K$
\\
$T^{\ }_{AF}$
&  $0.2074 J$    & $0.1769 J$          & $0.225 J$        
\\
$J^{\ }_{XY}/J$
& $5.7\times 10^{-4}$
& $5.3\times 10^{-4}$
& 
\\
$J_{DM}/J$
& $1.5\times 10^{-2}$
& 
& $7\times 10^{-2}$
\\
$(J'/J)^{\ }_{\mathrm{sw}}$
&$5\times 10^{-5}$& $10^{-8}$           & $5\times 10^{-5}$ 
\\
$2\pi\rho_s$
& $1.131 J $     & $1.15 J$            & $1.31 J$
\\
$c$ 
& $1.669 Ja$     & $1.669 Ja$          & $1.669 Ja$
\\
$\xi^{(2)}(T_{AF})$
& $105.6 a$      & $303 a $            & $69.77 a$          
\\
$\xi'$
&$215.81 a$& $1.548\times 10^{4} a$           & $207.197 a$ 
\\
\hline\hline
\end{tabular}
\caption{\label{tab: spin-1/2} 
Relevant parameters for some $S=1/2$ quasi-two-dimensional AF.
The intraplanar nearest-neighbor AF Heisenberg 
exchange coupling is $J$, 
while $J'$ is the AF Heisenberg coupling
between nearest-neighbor layers. 
The $XY$ exchange coupling is $J^{\ }_{XY}$, while
$J^{\ }_{DM}$ is the coupling of a Dzyaloshinsky-Moriya interaction.
The N\'eel temperature is $T_{AF}$.
The two-dimensional spin stiffness
$2\pi\rho^{\ }_s$ and 
spin-wave velocity $c$ are obtained either from large $S$ expansions
or from MC simulations.
The two-dimensional correlation length
$\xi^{(2)}(T^{\ }_{AF})$ is obtained from formula~(\ref{CHN-HN}).
Lengths are measured in units of the lattice spacing $a$.
}
\end{table}

\subsection{
Experimental discussion of the third scaling law
           }

What has been extensively studied experimentally is an 
inelastic neutron-scattering measurement that is
believed to yield an approximation to the instantaneous structure factor
\begin{eqnarray}
S^{\ }_{0}(\beta):=
\int\limits_{-\infty}^{+\infty} \frac{d\omega'}{2\pi}
S^{\ }_{0}(\pi,\pi,0,\omega';\beta).
\end{eqnarray}
Two ranges of temperatures have been studied in the literature. 
In the 1970s the chosen temperature range was a very narrow one 
about the AF transition temperature. The rationale for this choice was
to study the critical regime surrounding the ordering temperature.
In the 1990s the temperature range was broader and above the onset of three-dimensional
critical fluctuations.\cite{Keimer92}$^{-}$\cite{Leheny99}
The rationale was primarily to study two-dimensional
fluctuations associated with the classical renormalized regime of a
two-dimensional AF. In either cases, the experimental measurement of
$S^{\ }_{0}(\beta)$ is performed in arbitrary units, i.e., the overall scale
of $S^{\ }_{0}(\beta)$ is unknown. Since we are predicting a universal number 
when measured in some given units, we need to convert any measured number 
in arbitrary units $S^{(\mathrm{a.u})}_{0}(\beta)$
into a number in some chosen units. We do this by multiplying
$S^{(\mathrm{a.u})}_{0}(\beta)$
with a conversion factor,
\begin{eqnarray}
S^{\ }_{0}(\beta)\sim
S^{(\mathrm{a.u})}_{0}(\beta)
\times
\frac{S^{(\mathrm{MC})}_{0}(\beta^{*})}{S^{(\mathrm{a.u})}_{0}(\beta^{*})},
\end{eqnarray}
obtained by taking the ratio of the number $S^{(\mathrm{MC})}_{0}(\beta^{*})$
at inverse temperature $\beta^{*}$ computed within some scheme 
with the measured number in the same arbitrary units $S^{(\mathrm{a.u})}_{0}(\beta^{*})$
at the inverse temperature $\beta^{*}$.

We are only aware of the published MC computation by Kim and Troyer
in Ref.\ \onlinecite{Kim98} of the static structure factor at $(\pi,\pi)$
of a spin-1/2 AF on a square lattice.
We use this computation at temperatures such that the corresponding correlation length
is $3$ in units of the lattice spacing. These are high temperatures for which
the two-dimensional static structure factor should be a good approximation 
to the three-dimensional one. We are not aware of a published calculation of
the static structure factor using either MC or high-temperature series 
expansion for spin-1 or spin-5/2 AF on a square lattice that are needed to reinterpret
from our point of view the experiments from Refs.~\onlinecite{Nakajima95} or
Refs.~\onlinecite{Fulton94}, \onlinecite{Lee98}, and \onlinecite{Leheny99},
respectively. 
If we restrict ourselves to the experiments on quasi-two-dimensional 
spin-$1/2$ from Refs.\
\onlinecite{Birgeneau99}
\onlinecite{Greven95},
and
\onlinecite{Ronnow99},
we fail again to observe a signature of universality in $J'\beta S^{\ }_{0}(\beta)$
when measured at the temperatures
$T=337K,\ 278K,\ 18K$
that are 
$12K$, $21.5K$, and $1.5K$
above the corresponding ordering temperatures, respectively, as we find that
$J'\beta S^{\ }_{0}(\beta)$ takes the values
$1.17$,
$0.001$,
and
$0.077$,
respectively.

\section{
Conclusions
        }
\label{sec: Conclusions}

As we have seen, interesting scaling relations can be established between the 
interplane coupling $J'$ and observables of the underlying two-dimensional system
or the quasi-two-dimensional system.
The large-$N$ approximation is particularly well suited 
to compute the universal functions
obtained in the limit $J'\rightarrow 0$ at the N\'eel temperature
and derive the leading nonuniversal corrections.
The saddle-point approximation already leads to prediction in qualitative
agreement with Monte Carlo (MC) simulations.
To compare with existing experiments, we have included corrections of order $1/N$.
These corrections allow us to take into account the renormalization of 
the wave function as well as the renormalization of the interlayer coupling,
but will probably not affect the qualitative picture obtained in the saddle-point 
approximation.

The analysis of the quasi-two-dimensional model has revealed the existence of
different regimes at low temperatures, 
corresponding to the renormalized classical (RC), 
quantum critical (QC),
and quantum disordered (QD) regimes of the two-dimensional underlying system.
The universal constant obtained from the first and third scaling laws in the limit
$J'/J\to 0$ are strongly modified depending on the regime considered.
For the second scaling law, we did not find any distinction between the three regimes
in the saddle-point approximation. However, we expect different corrections for the
RC, QC, and QD regimes to the next order in the $1/N$ expansion. 
While the RC result will probably not be strongly affected by renormalizations,
the constant obtained in the quantum regimes might be modified more consequently 
by higher orders of the $1/N$ expansion.

The first scaling law in the RC qualitatively agrees with the MC simulations 
of Yasuda {\it et al.} in Ref.~\onlinecite{Yasuda05} for small $J'/J<0.1$.
Small variations of the N\'eel temperature
are exponentiated in the correlation length.
As a consequence, the first scaling law is very sensitive
to small errors in the expression for the N\'eel temperature,
so that a quantitative agreement between numerical 
simulations and our results seems difficult to achieve, 
even going beyond the saddle-point approximation. %

Recent MC simulations of quasi-two-dimensional systems 
in the RC, QC, and QD regimes have been performed by Yao and Sandvik
in order to test the scaling relations described in the present paper.%
~\cite{Yao06}
Their numerical results share many qualitative properties with
our predictions.

It is instructive to view our first scaling law against
the mean-field estimate for the N\'eel temperature~\cite{Villain77,Chakravarty88}
obtained by balancing the gain in energy derived from aligning planar spin $S$ 
with a mean-field magnetic field $SJ'[\xi^{(2)}(T)]^2$ parallel to
the stacking direction with the thermal energy $T$. This would give the
naive estimate
\begin{eqnarray}
\frac{J'}{J}\left(\frac{\xi^{(2)}(T^{MF}_{AF})}{a}\right)^{2}\sim
\frac{T^{MF}_{AF}}{J}.
\label{eq: Villain+Loveluck concl I}
\end{eqnarray}
Instead, we can improve this estimate by including 
two types of logarithmic corrections. We can replace the bare
ratio $J'/J$ by $(J'/J)^{\ }_{\mathrm{sw}}$
and we can multiply the right-hand side of Eq.~(\ref{eq: Villain+Loveluck concl I})
by logarithmic corrections in the
two-dimensional correlation length $\xi^{(2)}(T)$ measured in units of the
lattice spacing $a$.  The N\'eel temperature is then the solution to
\begin{eqnarray}
\left(\frac{J'}{J}\right)^{\ }_{\mathrm{sw}}
\left(\frac{\xi^{(2)}(T^{\ }_{AF})}{a}\right)^{2}\sim
\frac{T^{\ }_{AF}}{J}\ln^{q}\left(\frac{\xi^{(2)}(T^{\ }_{AF})}{a}\right)
\label{eq: Villain+Loveluck concl}
\end{eqnarray}
with the exponent $q$ measuring the strength of the logarithmic corrections.
Note that while Eq.~(\ref{eq: Villain+Loveluck concl} gives the
correct scaling, it may have multiplicative
nonuniversal corrections as exemplified by the presence of the lattice
constant $a$ on the left-hand side.
For the case of the $O(N)$ symmetry classes with $N>2$, we have in fact shown
the existence of logarithmic corrections with
a quantum origin that are induced by 
the substitution $J'/J\to(J'/J)^{\ }_{\mathrm{sw}}$
and of logarithmic corrections with a classical origin through the 
validity of Eq.~(\ref{eq: Villain+Loveluck concl}) 
with the extrapolation
\begin{eqnarray}
q=\frac{N-1}{N-2}
\end{eqnarray}
of our leading ($q=1$)
and first subleading calculation ($q=1+2/N$) in the $1/N$ expansion.
It would be interesting to investigate
the symmetry classes $N=1$ (Ising) and $N=2$ ($XY$)
from this point of view.

The most important limitation to experimental observations
of the scaling laws proposed here are symmetry crossovers.
Indeed, most of the candidates for quasi-two-dimensional
antiferromagnets have a spin anisotropy which is larger than
the anisotropy in space. 
The anisotropy in the spin space leads to a crossover from
an $O(3)$ Heisenberg model to an Ising or $XY$ Heisenberg model
which dominates over the dimensional crossover.
The organic compounds (5MAP)$_2$CuBr$_4$ 
and (5CAP)$_2$CuBr$_4$ seem to be promising candidates
for a magnet in which the dimensional crossover dominates 
over the symmetry crossover.%
~\cite{Woodward02}
It would therefore be interesting to have independent estimates
of the interlayer coupling constant $J'/J$ from the spin-wave 
dispersion on the one hand as well as from a measurement of the 
two-dimensional correlation length on the other hand for this
class of compounds. 

\section*{Acknowledgments}
We would like to thank Daoxin Yao and Anders Sandvik
for showing us their numerical simulations before publication
and Henrik R\o nnow for useful discussions.
M.B.H. was supported by U. S. DOE contract W-7405-ENG-36.

\appendix

\section{
Quasi-one-dimensional Ising model
        }
\label{app: Quasi-one-dimensional Ising model}

The first universal relation proposed in Refs.~\onlinecite{Yasuda05} 
and~\onlinecite{Hastings06} can be established exactly for the strongly 
anisotropic two-dimensional ferromagnetic Ising model
using the Onsager solution (see, for example, Ref.~\onlinecite{Izyumov}). 

Similarly to the two-dimensional Heisenberg AF, 
the one-dimensional Ising chain does not order at any finite temperature. 
Using the matrix formalism, the spin susceptibility of the Ising chain can 
be computed exactly,
\begin{subequations}
\begin{equation}
\chi_{s}^{(1)}(\beta_c)
=
\frac{\beta}{\tanh (J\beta)}\frac{e^{1/\xi}+1}{e^{1/\xi}-1},
\end{equation}
where the correlation length is
\begin{equation}
\xi = -\frac{1}{\ln\tanh (J\beta)}.
\end{equation}
\end{subequations}
The two-dimensional Ising model has a nonvanishing transition temperature, 
which can be obtained as the solution to
\begin{subequations}
\begin{equation}
\sinh\big(2\beta_c J\big)\sinh\big(2\beta_c J'\big)=1,
\label{app eq: implicit critical temperature}
\end{equation}
where $J$ and $J'$ are the nearest-neighbor exchange couplings 
along and between the Ising chains, respectively.
The solution of this equation is approximately
\begin{equation}
\beta_c
\approx 
\frac{1}{2J}\ln \frac{2J}{J'}
-\frac{1}{2J}\ln \ln \frac{2J}{J'}.
\label{app eq: approximate critical temperature}
\end{equation}
\end{subequations}
Using these results, one can compute
the first scaling law,
\begin{equation}
J'\chi^{(1)}_{s}(\beta^{\ }_c) = 
\mathcal{F}^{\ }_1(J\beta^{\ }_c),
\end{equation}
where the scaling function $\mathcal{F}^{\ }_1$ can be evaluated exactly. 
In the limit of small $J'$, we obtain
\begin{equation}
\mathcal{F}^{\ }_1(x)\approx 
1 
+ 
2e^{-2x},
\qquad 
x=J\beta^{\ }_c. 
\end{equation}
As a function of $J'/J$, the scaling relation reads
\begin{equation}
J'\chi^{(1)}_{s}(\beta^{\ }_c)
\approx 
1+\frac{J'}{J}\ln \frac{2J}{J'}.
\end{equation}
This gives a value $\zeta = 1$, which is half the 
coordination number expected in RPA. 

\section{
Two different estimates for \texorpdfstring{$T^{\ }_{AF}$}{Tc}
}
\label{app:Two different estimates fo TAF}

The interlayer coupling $J'/J$ can be estimated from the N\'eel temperature,
provided the dependence of $T^{\ }_{AF}$ on $J'/J$ is known.
We will discuss here different approximations giving this dependence. 

In the RPA from Refs.~\onlinecite{Scalapino} and \onlinecite{Schultz96}, 
the effect of the interlayer coupling is encoded by
an effective magnetic field proportional to the staggered magnetization 
in the adjacent layers that is multiplied by $J'$. 
The value of the staggered magnetization is determined self-consistently. 
The RPA staggered spin susceptibility is found to be
\begin{equation}
\chi^{\mathrm{(RPA)}}_s(T) =
\frac{\chi^{(2)}_s(T)}{1-z J'\chi^{(2)}_s(T)},
\end{equation}
where $z$ is the coordination number of a single layer.
At the N\'eel temperature, the spin susceptibility diverges.
This condition leads to the following equation for the N\'eel temperature 
in the RPA
\begin{equation}
z J' \chi^{(2)}_s(T^{\mathrm{RPA}}_{AF})=1.
\end{equation}
This equation was the motivation for the first scaling law.
It turns out that the quality of this approximation is
ensured by the first scaling law.

Another estimate for the N\'eel temperature
in the quasi-two-dimensional Heisenberg AF 
can be obtained by comparing the thermal energy $T^{\ }_{AF}$
with the interaction energy between ordered spins
in adjacent layers.~\cite{Chakravarty88} 
This argument leads to Eq.~(\ref{eq: cond for Neel temp from NRJ}),
\begin{equation}
zJ' 
\left(\frac{M}{M^{\ }_{0}}\right)^2
\left(\frac{\xi^{(2)}(T^{\ }_{AF})}{a}\right)^2 \simeq
T^{\ }_{AF}
\label{app eq: cond for Neel temp from NRJ}
\end{equation} 
where $M/M^{\ }_0\simeq 0.36$ 
is the reduction in the staggered magnetization at $T=0$
due to the two-dimensional spin fluctuations at length scales 
shorter than $\xi^{(2)}(T)$. A similar equation has been derived 
for quasi-one-dimensional systems by Villain and Loveluck
in Ref.~\onlinecite{Villain77} 
based on a real-space decimation argument.
Equation~(\ref{app eq: cond for Neel temp from NRJ}) 
is in contradiction with the first scaling law
that leads to Eqs.~(\ref{eq: ratio xi' / xi^{2d}, RC})
and (\ref{eq: ratio xi' / xi^{2d}, RC 1/N}).

An estimate of the interlayer coupling $J'/J$ from 
the N\'eel temperature and from the two-dimensional correlation
length extrapolated down to $T^{\ }_{AF}$
can be obtained using Eq.~(\ref{eq: ratio xi' / xi^{2d}, RC 1/N}).
Inserting $\xi'$ from  Eq.~(\ref{eq: xi prime from spin-wave J'/J})
in Eq.~(\ref{eq: ratio xi' / xi^{2d}, RC 1/N}) leads to
\begin{equation}
\left(\frac{J'}{J}\right)^{\ }_{\mathrm{sw}}=
K
\left(\frac{\xi^{(2)}(T^{\ }_{AF})}{a}\right)^{-2}
\left(\frac{4\pi\rho^{\ }_s}{(N-2)T^{\ }_{AF}}\right)^{1/(N-2)},
\label{app eq: cond for Neel temp from 1st scaling law}
\end{equation}
where $K=\Xi^2/\big(1+\frac{1.069}{N}\big)$. 
For instance, using 
$2\pi\rho^{\ }_s=1.131 J$, 
$\xi^{(2)}(T^{\ }_{AF})\simeq 3.3 a$,
and $K=0.937$ for $N=3$ leads to 
\begin{equation}
(J'/J)^{\ }_{\mathrm{sw}}\simeq 0.32
\end{equation}
for the organic compounds A$_2$CuX$_4$ with $T^{\ }_{AF}=0.6 J$.
This result is bigger than the estimate $J'/J\simeq 0.24$ 
proposed by Yasuda {\it et al.} in Ref.~\onlinecite{Yasuda05}
using their phenomenological formula based on MC simulations,
which itself is larger than the estimate $J'/J\simeq 0.08$
from Woodward {\it et al.} in Ref.~\onlinecite{Woodward02} based on 
Eq.~(\ref{eq: cond for Neel temp from NRJ}).
However, in view of the relative large $J'/J$, we expect large corrections
to our scaling laws.
 
As compared to the mean-field result%
~(\ref{app eq: cond for Neel temp from NRJ}), 
Eq.~(\ref{app eq: cond for Neel temp from 1st scaling law})
contains logarithmic corrections. Indeed, using
\begin{equation}
\frac{2\pi\rho^{\ }_s}{(N-2)T^{\ }_{AF}} \simeq
\ln\left(\frac{\xi^{(2)}}{a}\right)
+
\mathcal{O}\left[\ln\ln\left(\frac{\xi^{(2)}}{a}\right)\right],
\end{equation}
we can rewrite Eq.~(\ref{app eq: cond for Neel temp from 1st scaling law}) as
\begin{equation}
\left(\frac{J'}{J}\right)^{\ }_{\mathrm{sw}} 
\left(\frac{\xi^{(2)}(T^{\ }_{AF})}{a}\right)^2\sim
\frac{T^{\ }_{AF}}{J}\ln^q\left(\frac{\xi^{(2)}(T_{AF})}{a}\right),
\label{app eq: log correction to mean-field Neel temp}
\end{equation}
where 
\begin{equation}
q=\frac{N-1}{N-2}.
\end{equation}
In particular, $q=1$ in the $N=\infty$ approximation, while for $N=3$ we find $q=2$.
The same value $q=2$ can be obtained in the classical case,
using the results of Br\'ezin and Zinn-Justin.~\cite{Brezin76}
Indeed, they found in a $2+\epsilon$ expansion up to two loops, 
\begin{subequations}
\begin{equation}
\chi_s^{(2)}(T)\sim T^3 \exp(4\pi J/T),
\end{equation}
for the staggered spin susceptibility and 
\begin{equation}
\xi^{(2)}(T)\sim T \exp(2\pi J/T) 
\end{equation}
\end{subequations}
for the correlation length.
Equation~(\ref{app eq: log correction to mean-field Neel temp})
with $q=2$ then follows from the first scaling law~(\ref{eq: first scaling law}).



\begin{thebibliography}{99}
\bibitem{footnote: Z'}
The multiplicative renormalization $\mathcal{Z}^{\prime}$
is closely related but not identical to the 
renormalization $Z^{\prime}$ introduced in Eq.~(5c) of 
Ref.~\onlinecite{Hastings06}.
The dimensionless interlayer spacing $\xi^{\prime}/a$
is related to the renormalized interlayer coupling
$\alpha_{c}=\big(\xi'/a\big)^{-2}$ introduced in Eq.~(76)
from Ref.~\onlinecite{Irkhin97}. It is determined from
computing order by order in a $1/N$ expansion the 
corrections to Eq.~(\ref{eq: integral form of G0 c})
in the limit $p^{\ }_{z}\to0$ and at the N\'eel temperature
($\lambda^{\ }_{*}=0$).
\bibitem{footnote: nnn nnnn etc}
Action~(\ref{eq: S = S1})
is not the most general one allowed by symmetry.
For example, integration over fast modes also induces 
longer range interlayer interactions
$$
\boldsymbol{\sigma}^{\ }_j\cdot\boldsymbol{\sigma}^{\ }_{j+1}
\to
\boldsymbol{\sigma}^{\ }_j\cdot\boldsymbol{\sigma}^{\ }_{j+1}
+
\sum_{n=2,3,\cdots}
\kappa^{\ }_{n}
\boldsymbol{\sigma}^{\ }_j\cdot\boldsymbol{\sigma}^{\ }_{j+n}.
$$
The dimensionless constants $\kappa^{\ }_{n}$
are of order $J'/J$ or higher and can be ignored in the limit
$J'/J\to0$ without loss of generality (unless specified).
\bibitem{footnote: Fourier conv}
The convention
$
\int_x\equiv\int_{0}^{\beta}d\tau\int d^{2}\boldsymbol{r}
$
with
$kx\equiv\boldsymbol{k}\cdot\boldsymbol{r}-\omega\tau$
and
$k\equiv(\omega/c,\boldsymbol{k})$
in Eq.~(\ref{eq: def Fourier trsf})
as well as the convention
$
\int_{\boldsymbol{p},p^{\ }_{z}}\equiv
\int\frac{d^2p}{(2\pi)^2}
\int_{0}^{2\pi}\frac{dp_z}{2\pi}
$
in Eq.~(\ref{eq: integral form of G0})
are used.
Observe that the two-dimensional susceptibility
(\ref{eq: def 2-dim iso  static spin susc})
follows from Eq.~(\ref{eq: def Fourier trsf})
by setting $\alpha=0$.
\bibitem{footnote: Z}
The multiplicative renormalization $\mathcal{Z}$
is closely related but not identical to the 
renormalization $Z^{\ }_{h}$ introduced in Eq.~(4b) of 
Ref.~\onlinecite{Hastings06}.
\end{thebibliography}
\end{document}